\documentclass[12pt]{article}
\usepackage{a4}
\usepackage{graphicx}
\usepackage{rotating}
\usepackage{hyperref}
\usepackage{xspace}
\usepackage[draft]{fixme}
\usepackage{amsmath}
\usepackage{numprint}
\usepackage{subfigure} 
\usepackage[T1]{fontenc}
\usepackage[latin1]{inputenc}
\usepackage{authblk}

\def\d{\ensuremath{\mathrm{d}}}
\def\emc{\texttt{emc2}\xspace}
\def\phb{\texttt{phb}\xspace}
\def\maps{\texttt{maps}\xspace}
\def\pollmach{\texttt{pollmach}\xspace}
\def\rsv{\texttt{runstruct\_vasp}\xspace}
\def\ezvasp{\texttt{ezvasp}\xspace}
\def\gnuplot{\texttt{gnuplot}\xspace}
\def\ch#1{\ensuremath{{\rm#1}}}
\title{Calculating phase diagrams with ATAT}
\author{Martin B\"aker}
\affil{Technische Universit\"at Braunschweig, 38106 Braunschweig,
  Germany, martin.baeker@tu-bs.de}
\begin{document}
\maketitle
\section{Introduction}
This document is a short and informal tutorial on some aspects of calculating phase
diagrams with the ATAT-tools \emc and \phb and on creating cluster
expansions with \maps. It is neither complete, nor in any way an
official document, but mainly a set of collected notes I took during
experimentation with ATAT. I assume that you have read at least
\cite{Walle2002} and have some basic idea about ATAT and what it does
(but may be stymied by the large number of options and program
variants). For the final sections on phonon and electron
contributions, reading \cite{Walle2009} is also helpful. Note that all
programs in the ATAT-toolkit have a \texttt{-h} option which gives a
detailed explanation of (most) parameters you can set and of input and
output files.

\section{Chemical potential}

The main variable to govern the composition of the alloy is the
chemical potential. Since this is a semi-grand canonical ensemble (SGCE), the
number of atoms is fixed (if vacancies are present, they are counted
as an atom species); each lattice site contains one atom. 

A change of concentration thus means replacing atoms of one sort with
one of the other. (For simplicity, we are looking at two species
only.) If we change one atom, we have (here and in the following, I
use $F$ for the free energy and do not consider $G$)
\cite[8.3]{frenkel2001understanding}
\begin{equation}
\frac{\d F}{\d n} = \mu_A - \mu_B =\mu
\end{equation}
since we change one atom from A to B. If the chemical potential is
defined wrt the concentrations, replace $n$ by the concentration $x$. 
Note that in ATAT (at least in the binary version), the chemical
potential used is the difference between the two species, so there is
only one value.

In the semi-grand canonical ensemble, the chemical potential is an
external control variable (like the pressure in the NPT ensemble). So
it is defined by the change in the free energy but it is nevertheless
prescribed externally. The chemical potential can be understood as the
thermodynamic force trying to ``push'' an atomic species into the
system. A chemical potential of~$0$ means that there is no external
driving force to replace one species with the other. 

In ATAT,
there are two conventions for the chemical
potential: In the output, it is given as a physical quantity in the
units used (depending, among others, on the value of $k_B$ given with
\texttt{-k}); in the input of \emc, the chemical potential is normalized so
that the region where the first phase (from \texttt{gs\_str.out}) is
stable corresponds to $\mu\in[0,1]$, the second phase is stable in the region
$\mu\in[1,2]$ and so on. (The disordered phase has negative
$\mu$). Note that the normalization is not done in this way if there
are only two or one ground states (see also section~\ref{sec:phasesep}).

Therefore, if you call \texttt{emc2} with\\
\texttt{emc2 -gs=1 -mu0=1.5 -mu1=0.5 -dmu=0.04 \ldots}\\
you start at a value of $1.5$ that perfectly stabilizes the second
phase (ground state number 1) and do calculations down to a value of
$\mu$ where the first phase (ground state number 0) is stable. If the
specified $\mu$-value does not correspond to the specified ground
state, you may get a warning that the chemical potential does not stabilize
this ground state. 

\section{Phase diagrams}

Phase diagrams can be drawn with the chemical potential instead of the
concentration as variable. In a two-phase region, because of the
standard common-tangent construction, the chemical potential is
constant. Therefore, the two-phase region collapses to a line in a
$\mu$-$T$ phase diagram. The advantage of using $\mu$ (and thus the SGCE) instead of $x$
is that  only one phase is stable at any value
of $\mu$ (unless we are \emph{exactly} at a $\mu$-value that
corresponds to a 2-phase region). Therefore, in a Monte-Carlo
simulation, the system should never separate into different phases
\cite[p. 24]{Morgan2005}. 

To calculate phase boundaries, the ``thermodynamic function''
\cite{Walle2002,Morgan2005} 
\begin{align}
\phi(\beta, \mu) &= E -TS- \mu x = -\frac{1}{\beta N} \ln\sum_i\left( 
\exp(-\beta N ( E_i - \mu x_i))
\right)\\
&= F- \mu x
\end{align}
is used.

This can be calculated for each phase -- the thermodynamically stable
phase will have the lower value of $\phi$; on a phase boundary, the
values for both phases are equal. Therefore, phase boundaries between
phase $\alpha$ and $\gamma$ are
defined by
\begin{equation}
\Lambda^{\alpha,\gamma} = \{
(\beta, \mu): \phi^\alpha(\beta, \mu) = \phi^\gamma(\beta, \mu)
\}\,.
\end{equation}

Since the chemical potential is conjugate to the concentration in a
SGCE, the concentration~$x$ in each phase can be calculated from
\begin{equation}
x^\alpha = - \frac{\partial \phi^\alpha(\beta, \mu)}{\partial \mu}\,.
\end{equation}
Since cluster expansions have their origin in an analogy to the Ising
model, the concentration runs from $-1$ (pure A) to $+1$ (pure B). To
get physical concentrations, use $(x+1)/2$. 

To calculate the phase boundary, $\phi$ has to be calculated, but with
MC simulations, only differences (total differentials) can be
calculated. The method can still be employed by starting at a point
where $\phi$ is known, for example from a low-temperature
expansion. This is what the program \texttt{phb} does when the option
\texttt{-ltep} is used.

The program \texttt{emc2} also calculates $\phi$; there, an LTE or HTE
is used if no initial value of $\phi$ (option \texttt{-phi0}) is
given.

\section{Tracking phase boundaries}
The basic strategy to follow the phase boundary is from
\cite[eq. (29)]{Walle2002} or \cite{walle}:
\begin{equation}
\frac{\d \mu}{\d \beta}= \frac{E^\gamma - E^\alpha}{\beta(x^\gamma -
  x^\alpha)} -\frac{\mu}{\beta}\,. \label{eq:trackphb}
\end{equation}
So we can calculate the change of $\mu$ with temperature from a given
point. 
So  the calculation proceeds by starting at a known
$(T,\mu)$-point and goes on from there, incrementing $\mu$ in finite
steps of $\beta$ (or the temperature).

The parameter \texttt{-dT} in \phb thus affects the
precision of following the boundary: Making it smaller means that it
is easier to follow a curved boundary. 
 
To check how this works, we can use the example file provided in the
\texttt{mc}-folder of ATAT:
\begin{verbatim}
T          mu        x1         x2         E1           E2
240  -0.0775028  -0.986175  -0.502882  -0.0495888  -0.0501994  
250  -0.0774447  -0.982053  -0.503663  -0.0493333  -0.0501169	
\end{verbatim}
The corresponding beta-values are
$1./(240\cdot\numprint{8.617e-5})  = 48.35403$ and 
$1./(250\cdot\numprint{8.617e-5})  = 46.41987$. 
Thus, $\delta\beta$ is 
$48.35403 -46.41987 = 1.93416$.
With these values, the left-hand side of the above equation is
\begin{equation*}
   (-0.0775028 + 0.0774447 ) / 1.93416  = \numprint{-3.0039e-5} \,.
\end{equation*} 
In the current version of ATAT (5/10/18), the last two columns are
actually $E-\mu x$, so the second term of eq.~\ref{eq:trackphb} is
already included in the ``energy''. The rhs thus becomes
\begin{equation*}
((-0.0495888 + 0.0501994) / (-0.986175 + 0.502882)) / 48.35402 = \numprint{-2.6128E-5}
\end{equation*}
which is close to the lhs (although not exactly identical).

A look into the source code of \texttt{phb.c++} shows in line 416:
\begin{verbatim}
      mu+=1.5*dmu-0.5*old_dmu;
\end{verbatim}
so the discretization is not as simple as I assumed here since it uses
the previous value of \texttt{dmu} as well. Nevertheless, the results are close
enough to see that the calculation of $\mu$ does in principle proceed as
explained here.

\section{The simplest possible example}\label{sec:phasesep}

%
%
We now try to perform a simple phase diagram calculation from scratch.
Use the following settings to create a binary system where the two
species sit on a simple cubic lattice and separate:\footnote{The
  atoms are called Al and Ni but have nothing to do with the
  materials, this is just an example.}

\begin{verbatim}
lat.in
3.5 3.5 3.5 90 90 90
1. 0 0
0 1 0
0 0 1
0 0 0 Ni, Al
\end{verbatim}

\begin{verbatim}
gs_str.out
3.500000 0.000000 0.000000
0.000000 3.500000 0.000000
0.000000 0.000000 3.500000
1. 0 0
0 1. 0
0 0 1.
1.000000 1.000000 1.000000 Ni
end

3.500000 0.000000 0.000000
0.000000 3.500000 0.000000
0.000000 0.000000 3.500000
1. 0 0
0 1. 0
0 0 1.
1.000000 1.000000 1.000000 Al
end
\end{verbatim}

These two set up a simple cubic lattice (in \texttt{lat.in}) and the two ground-state
phases (pure ``Ni'' and ``Al'' which of course would not form a simple
cubic lattice). Note that the actual value of the lattice
constant is totally irrelevant here since ATAT only looks at cluster
configurations.

The clusters are defined in
\begin{verbatim}
clusters.out:
1
0.000000
0

1
0.000000
1
1.000000 1.000000 1.000000

6
3.5
2
1.000000 1.000000 1.000000
1.000 1.0000 0.0000
\end{verbatim}
The first term is a constant term (with no atoms considered), the
second is the  single atom cluster (its expansion coefficient gives the energy difference between a
single Ni and a single Al atom), the third is the interaction of
nearest neighbours. 

Here are the coefficients to be used with the clusters:
\begin{verbatim}
eci.out
0.
0.
-1
\end{verbatim}
The only one that is really important is the third. It is $-1$, so
when considering two identical atoms (species both $+1$ or $-1$), the
energy contribution is negative, when considering non-identical atoms,
the energy is positive. Therefore, we should expect the system to
separate into two phases. 

A calculation using \phb shows that the chemical potential (as output
variable, see above) is zero at the phase boundary at 0~K, which
should be expected because of symmetry. The energy per atom is $-3$
(each atom has 6 nearest neighbours, but to avoid double-counting of
pair bonds, the energy per atom is $3\cdot(-1)$).

\begin{figure}
\includegraphics[width=7cm]{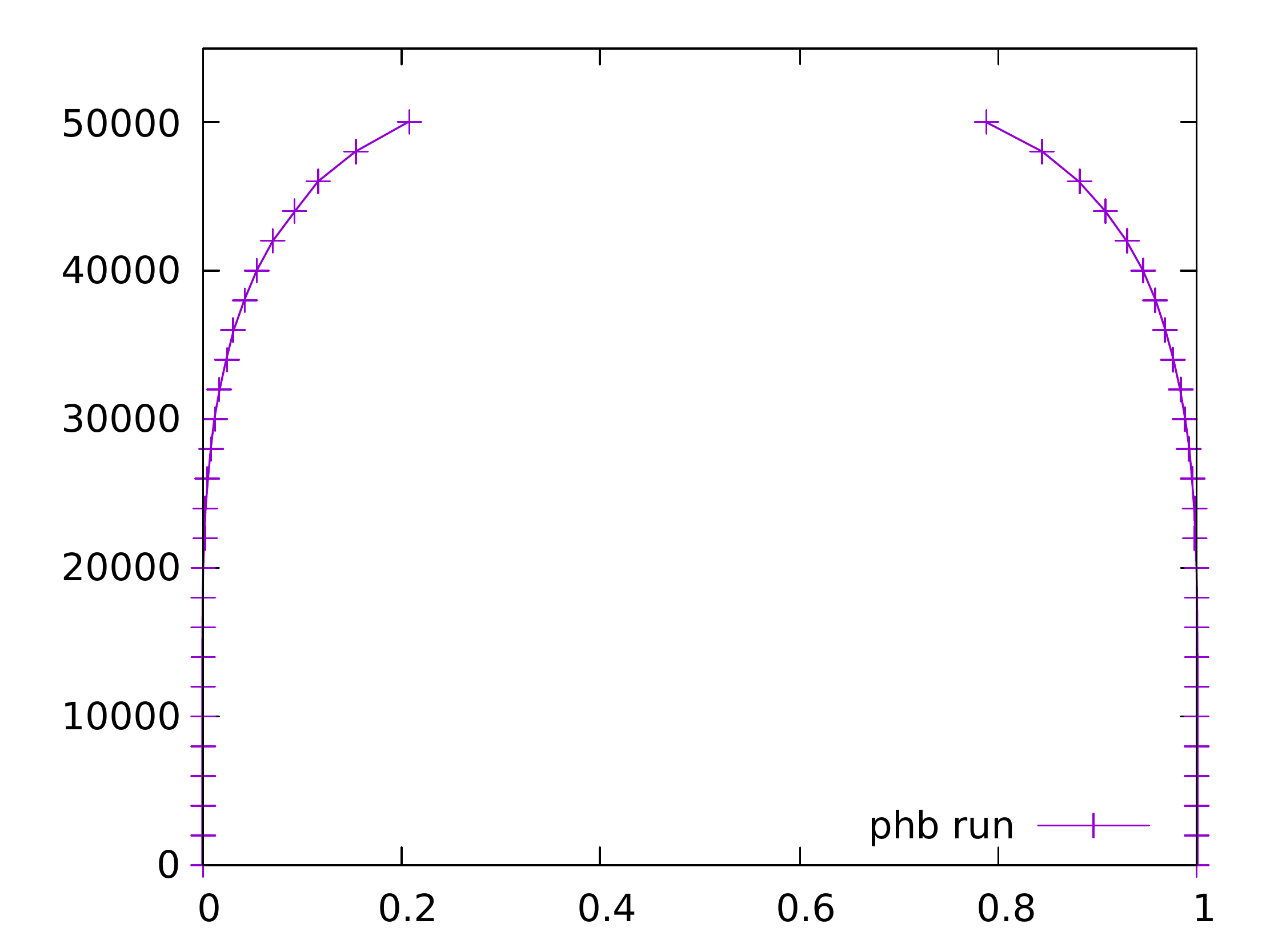}
\includegraphics[width=7cm]{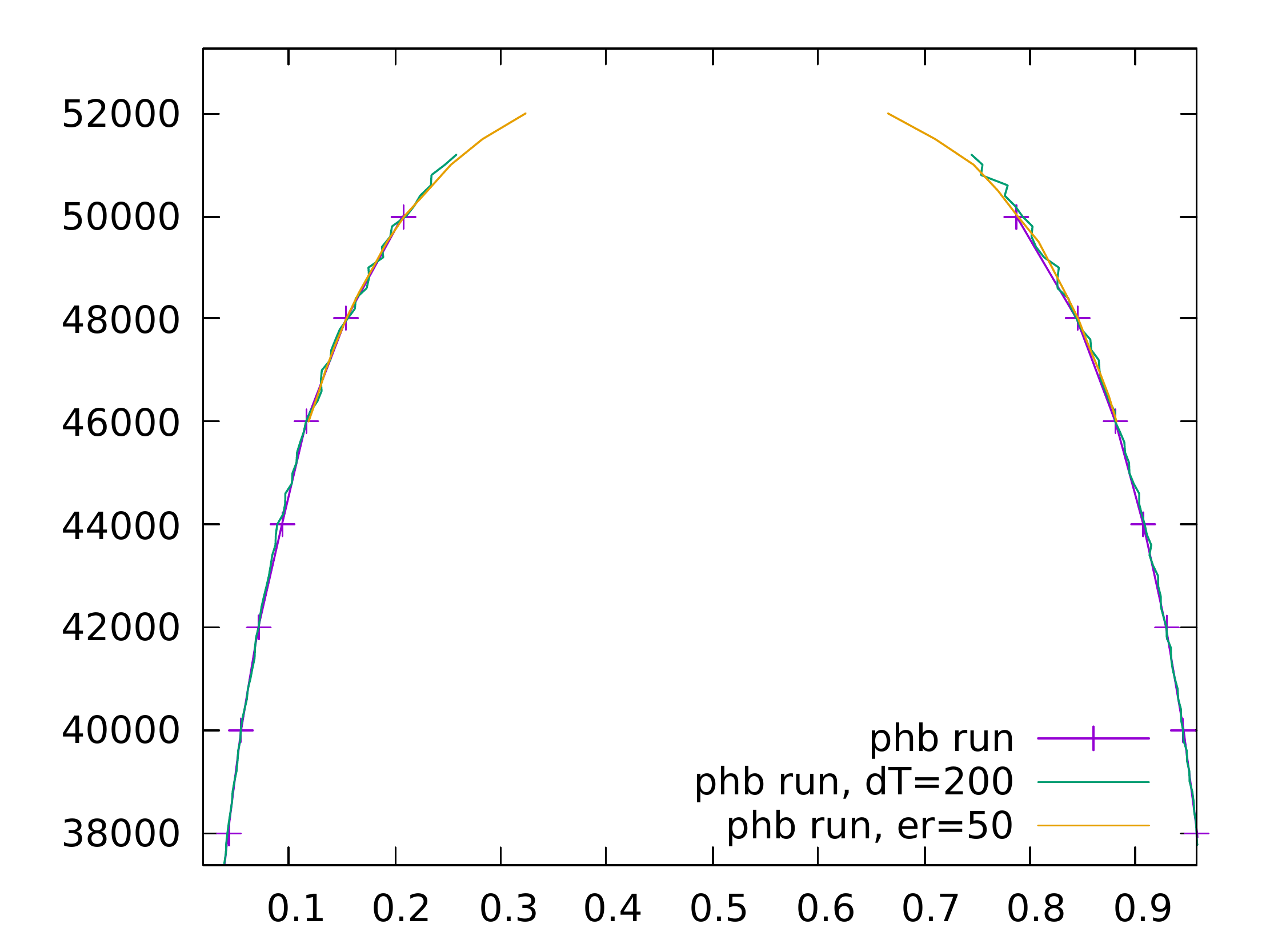} 
\caption{Phase diagram for the simple system of non-mixing
  species. Left: initial run with open gap. Right: Trying to close the
  gap by changing \texttt{dT} or \texttt{er}.
}
\label{fig:phaserunf2}
\end{figure}

To calculate the phase diagram, use (note that linebreaks in command lines
should of course not be typed in):
\begin{verbatim}
phb -gs1=0 -gs2=1 -dT=2000 -dx=1e-2 -er=20 -k=8.617e-5 
-ltep=5e-3 -o=ph01.out 
\end{verbatim}
The resulting phase diagram, fig.~\ref{fig:phaserunf2}, left, shows the miscibility gap.
You can plot it with \gnuplot \cite{gnuplot} using
\begin{verbatim}
plot [0:1][0:55000] "ph01.out" using (($3+1)/2):1 w lp,\ 
"ph01.out" using (($4+1)/2):1 w lp
\end{verbatim}

The calculation of \phb proceeds rapidly up to a temperature of 50000,
but then it stalls. The output from \phb shows that a huge number of MC
iterations would be required:
\begin{verbatim}
Phase 1 n_equil= 0 n_avg= 1280000
\end{verbatim}
This is an example of critical slowing down, i.\,e. the difficulty of
finding a new configuration that is statistically independent  from the
current one.  So unless you
are willing to wait for a very long time, there will be a gap
in your phase diagram at the top of the msicibility gap.

You can try to reduce the temperature-steps to make the gap smaller:
\begin{verbatim}
phb -gs1=0 -gs2=1 -dT=200 -dx=1e-2 -er=20 -k=8.617e-5
 -ltep=5e-3 -o=ph01-200.out
\end{verbatim}
The result is shown in fig.~\ref{fig:phaserunf2}, right.
Still, at a temperature of 51000, the number of equilibration steps
becomes rather large (72000). You can of course wait for this to
finish, it won't take too long.

One other way to deal with this that actually helps is to increase the
radius \texttt{er}:
\begin{verbatim}
phb -gs1=0 -gs2=1 -T=46000 -mu=2.9118e-5 -dT=500 -dx=1e-2 -er=50
 -k=8.617e-5 -ltep=5.e-3 -o=ph01-er50.out
\end{verbatim}
Here we start at a reasonable point for $T$ and $\mu$ (taken from the
previous run, of course the phase boundary should be exactly at
$\mu=0$ in this example) but use a larger cell. 
This run reaches a temperature of 52000 faster than the
other two. (After that, it also stalls) This might seem surprising
because usually critical slowing down
becomes worse when the simulation volume gets larger. I suspect (but
do not know) that this is because ATAT calculates the number of runs
needed and there is a trade-off between the needed number of runs and
the simulation volume. So it seems that smaller cells will not always
run faster. The resulting phase diagram is almost closed, but not
quite. The figure also shows that the run with the larger cell is
smoother as should be expected. (And a radius \texttt{er} of 20 is
really a bit small for a MC calculation.)

If this does not work, here is some advice from Axel van de Walle
[private comm.]:
\begin{quote} 
The way I usually handle this is by running \emc around the
expected top of the miscibility gap to explore where the gap may
start. \emc also has the option of running for a fixed number of steps
(\texttt{-n} option), which bounds the stalling time. When you have the top
point, you can usually just join it with the curves from phb and get a
decent-looking phase diagram.
\end{quote}

The idea behind this is that \emc runs at fixed value of $\mu$ so that
the system should always be in a single phase. (\emc actually contains
some checks on this, see below.) On varying $\mu$, you will then
``jump'' in the concentration from one phase to the other.
So, following this advice, run \emc as follows:\footnote{\emc writes
  several files: 
\texttt{ltedat.out},
\texttt{htedat.out},
\texttt{mfdat.out}, and
\texttt{mcsnapshot.out}. If you run several instances of \emc in the
same directory, be aware that these files will be overwritten. The
\emc-run itself is unaffected by this, but if you want to look at
configurations (\texttt{mcsnapshot.out}), it might be better to run
each \emc-instance in a separate directory.
}
\begin{verbatim}
emc2 -gs=0 -mu0=-0.5 -mu1=0. -dmu=0.1 -T0=2000 -T1=80000 -dT=2000 -keV
   -er=20 -dx=1.e-3 -o=emc-01.out > emc-01-run &
emc2 -gs=1 -mu0=0.5 -mu1=0. -dmu=0.1 -T0=2000 -T1=80000 -dT=2000 -keV
   -er=20 -dx=1.e-3 -o=emc-10.out > emc-10-run &
emc2 -gs=0 -mu0=-0.5 -mu1=0. -dmu=0.1 -T0=2000 -T1=80000 -dT=2000 -keV 
  -er=40 -dx=1.e-2 -o=emc-01-40.out > emc-01-40-run &
\end{verbatim}
Note that in this case, where we only have two ground states, the
$\mu$-parameter works differently. The help-page of \emc states:
\begin{quote}
  If there are only two ground states, the only correction performed is
  to shift mu so that mu=0 stabilizes a two-phase equilibrium between
  the two ground states.
\end{quote}
So in our case, the phase transition is at $\mu=0$ (no shift is
necessary) and not at $\mu=1$ (as you 
might expect, since in the general case, $\mu=1$ stabilizes the
boundary between phase 0 and 1). I did one run with a larger cell size
(and larger \texttt{dx}) to check that my cell is not too small.  
I also piped the output
to a file and sent the process to the background.

\begin{figure}
\includegraphics[width=7cm]{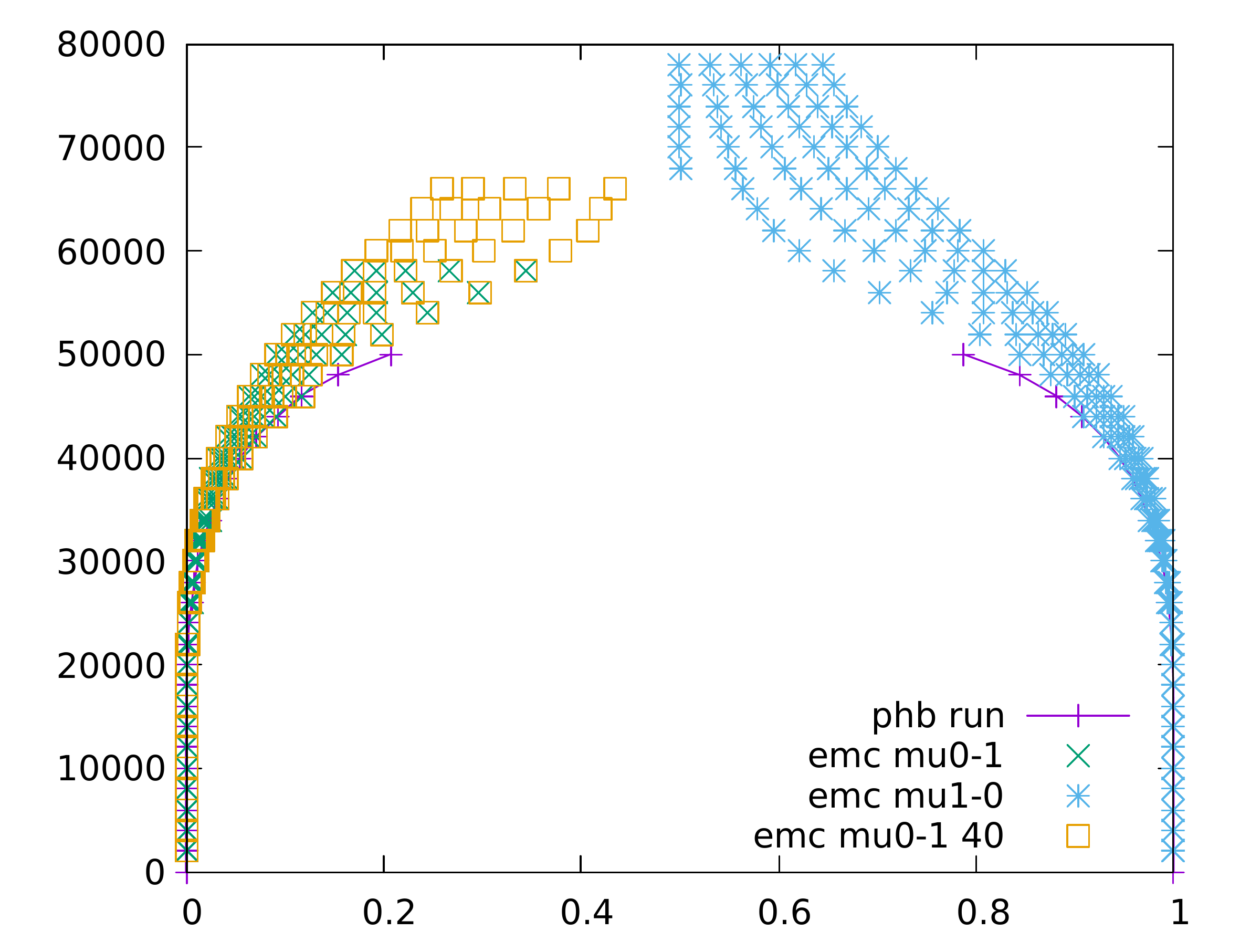}
\includegraphics[width=7cm]{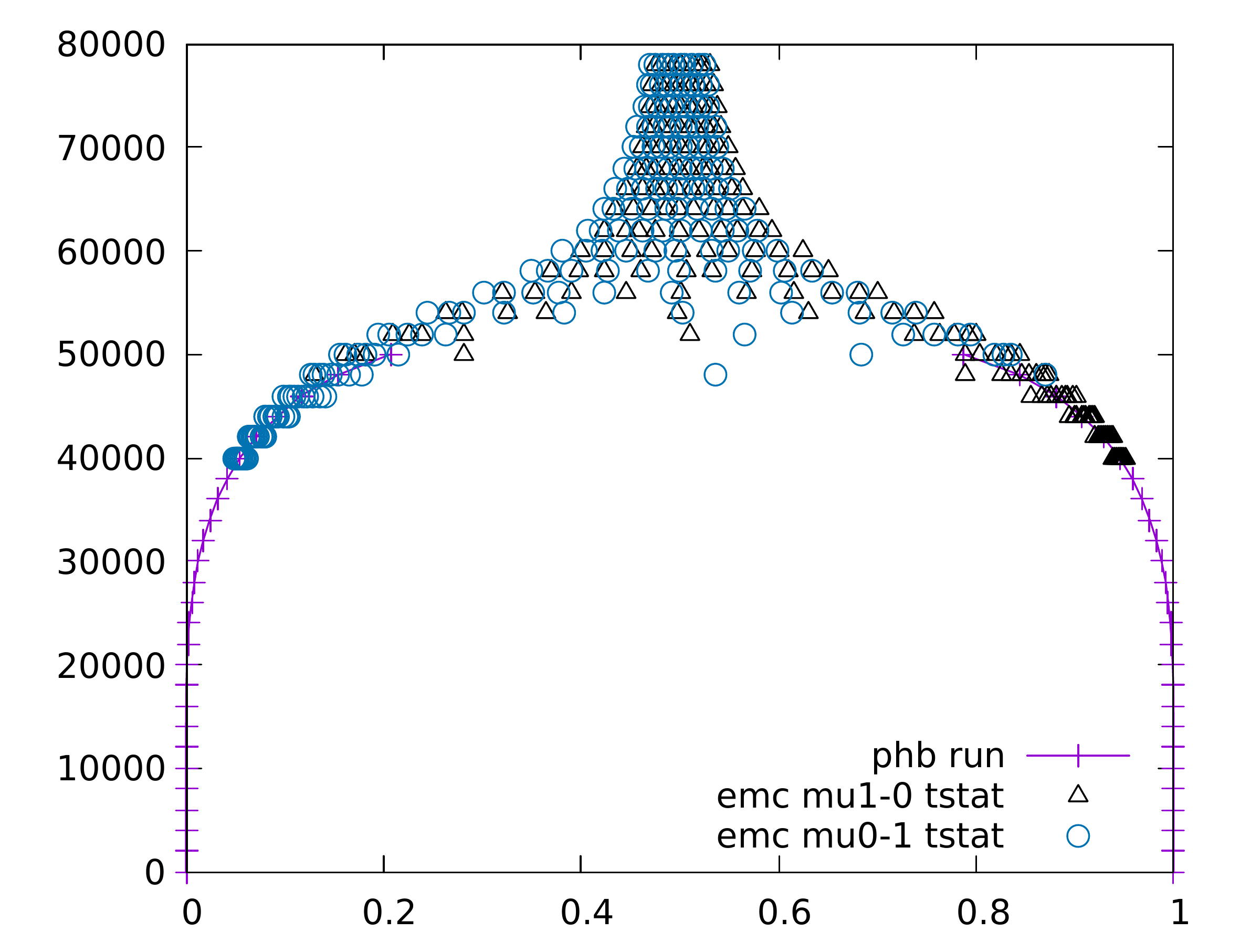}
\caption{Left: Running \emc on the simple system. The points looked at by
  \emc should all be in the one-phase region, so the boundary of these
  points gives a hint of the phase boundary. Right: Adding a run with \texttt{tstat=0}.
}
\label{fig:phaserunf3}
\end{figure}

We can now plot the $x$-$T$ plane of the states looked at by
\texttt{emc2} and compare them to the \texttt{phb}-diagram, see
fig.~\ref{fig:phaserunf3}, left. 
This plot is created using
\gnuplot:
\begin{verbatim}
set key bottom
plot "ph01.out" using (($3+1)/2):1 title "phb run" w lp ls 1,\
 "ph01.out" using (($4+1)/2):1 notitle w lp ls 1 ,\
 "emc-01.out"  using (($4+1)/2):1 title "emc mu0-1" ls 2,\
 "emc-10.out"  using (($4+1)/2):1 title "emc mu1-0" ls 3,\
 "emc-01-40.out"  using (($4+1)/2):1 title "emc mu0-1 40" ls 4
\end{verbatim}
I use columns 3 and 4 of the \phb-output (the two concentrations at
the phase boundary) and column 4 of the \emc-output (concentration at
the point currently looked at). Concentration values are converted
from $[-1,1]$ to $[0,1]$ as explained above. The output looks reasonable, although the
boundaries do not agree perfectly (deviation seems to be larger than
the required precision at higher temperatures).  This may be
due to a slight overshooting of \emc due to the hysteresis loop.

\begin{figure}
\includegraphics[width=7cm]{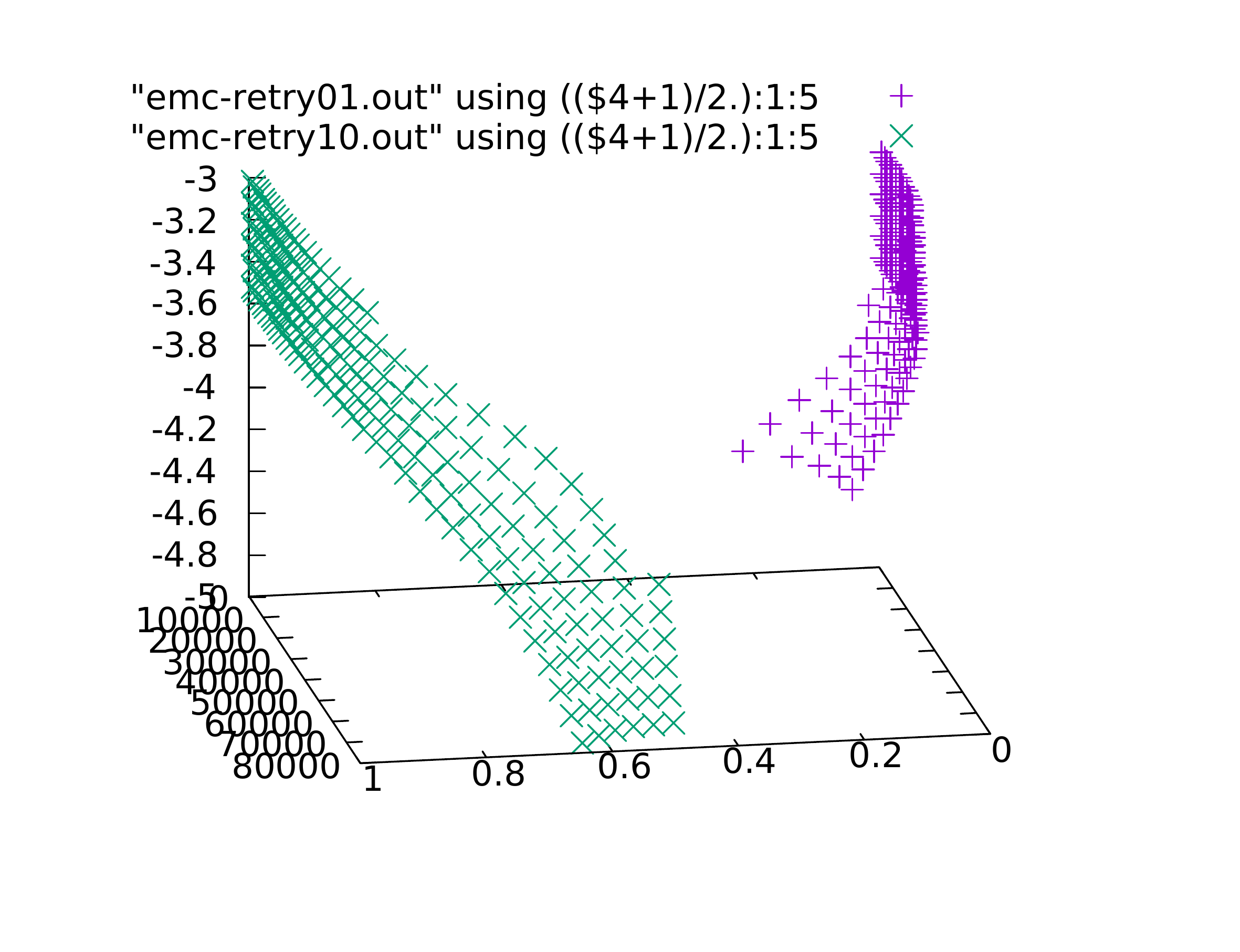}
\includegraphics[width=7cm]{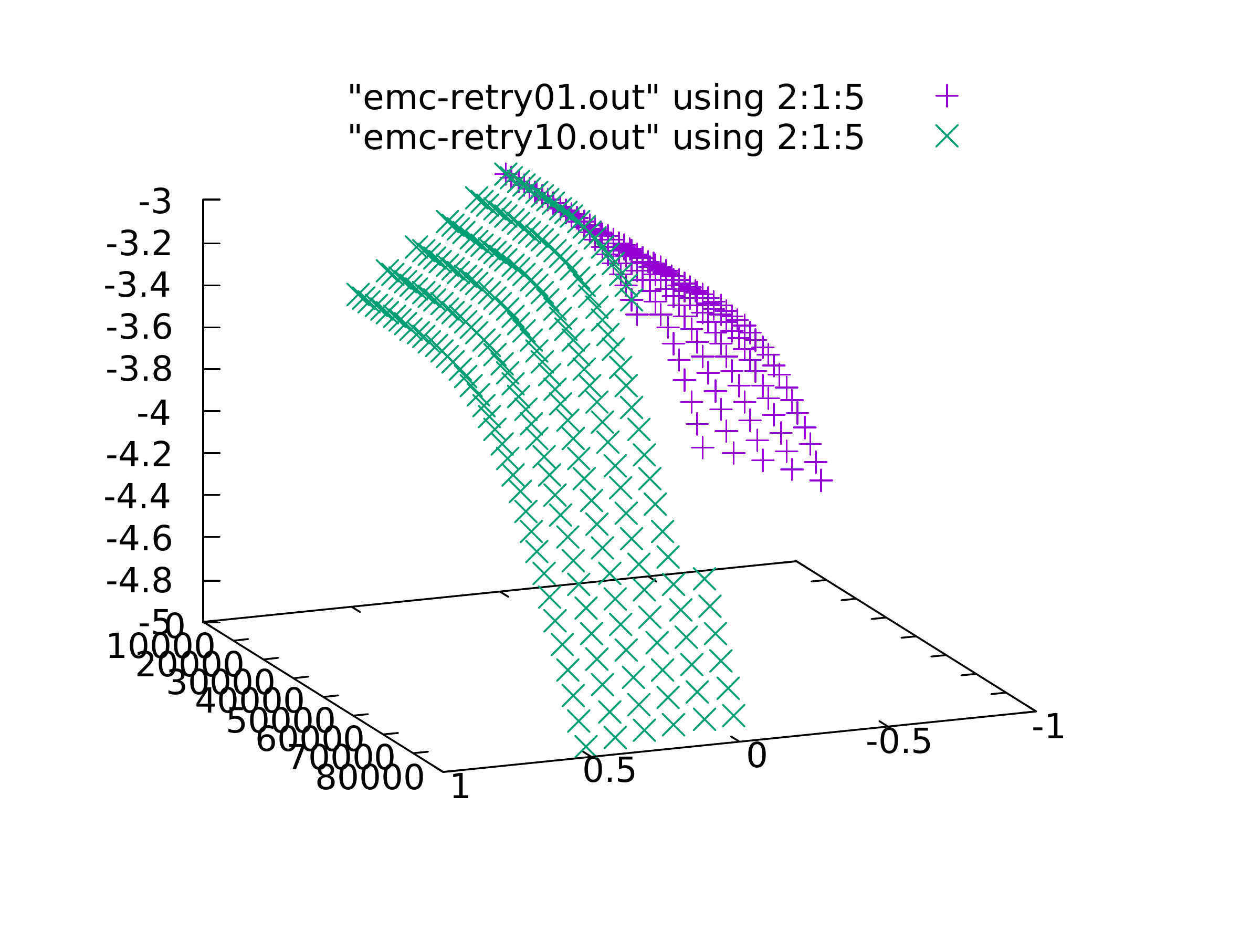}
\caption{Left:  potential $\phi$ vs. concentration and temperature; plotted with \texttt{gnuplot}
 Right: Plot of $\phi$ vs. $\mu$ and $T$ to show the phase transition
 at $\mu=0$.
}
\label{fig:phasesep2}
\end{figure}

You can also
plot the potential from \emc, for example 
\begin{verbatim}
splot  "emc-01.out" using (($4+1)/2):1:5,\
  "emc-12.out" using (($4+1)/2):1:5
\end{verbatim}
This plots the potential $F-\mu x$ versus the concentration and the
temperature. You expect to see a surface with gaps (in the
concentration region where two phases are in equilibrium) as shown in 
fig.~\ref{fig:phasesep2}.
To see the phase transition, we can plot $\phi$ vs. $\mu$ and $T$ as
shown on the right.  The two surface cross at
$\mu=0$ as expected.

The states in fig.~\ref{fig:phaserunf3} look reasonable, but they do
not follow the phase boundary very closely. You can estimate that the
top of the miscibility gap is somewhere between 52000 and 60000 which
is perhaps not precise enough.

One possibility to study this further is to switch off the checking of
phase transitions in \emc that causes \emc to stop at some point. This
is done with the option 
\texttt{tstat=0}.  Since this stalls after some time, I prescribe the
number of equilibrations and calculation steps. This is of course a
bit dangerous since you cannot rely on the statistics to be good
enough, so you have to experiment to see whether things work
out (for publishable results, you would surely use a larger cell and
more sweeps):\footnote{If you are
  patient enough, you can also specify \texttt{-dx}; at least for
  \texttt{-gs=0}, this finished after about a day of computing.}
\begin{verbatim}
emc2 -gs=0 -mu0=-0.1 -mu1=0.1 -dmu=0.02 -T0=40000 -T1=80000 -dT=2000 
  -keV -er=20 -n=5000 -eq=2000 -tstat=0 -o=emc-01-tstat.out
emc2 -gs=1 -mu0=0.1 -mu1=-0.1 -dmu=0.02 -T0=40000 -T1=80000 -dT=2000 
  -keV -er=20 -n=5000 -eq=2000 -tstat=0 -o=emc-10-tstat.out 
\end{verbatim}
The result  is
shown in fig.~\ref{fig:phaserunf3}, right. Note that -- as should be expected
-- some of the data points lie in the miscibility gap, but
there are some points almost exactly at $x=0$ that look like a good
candidate for the center of the miscibility gap. Look into the
corresponding output file from \emc to find their $\mu$- and $x$-values:
\begin{verbatim}
emc-01-tstat.out:
52000.000000	0.000000	-1.111069	0.131531	
54000.000000	0.000000	-0.899093	0.004387	
56000.000000	0.000000	-0.812865	-0.014910	
emc-10-tstat.out:
52000.000000	-0.000000	-1.108399	0.021828	
54000.000000	-0.000000	-0.905569	-0.006905	
56000.000000	-0.000000	-0.817109	0.001020
\end{verbatim}
The first concentration value (column 4) is a bit off, all others  are close to a
concentration of almost exactly 0. All their $\mu$-values
are 0. (This is the actual, physical $\mu$-value, not the value in the
ATAT-input convention. That it is exactly~0 is due to the symmetry in
this example.)

\begin{figure}
\includegraphics[width=7cm]{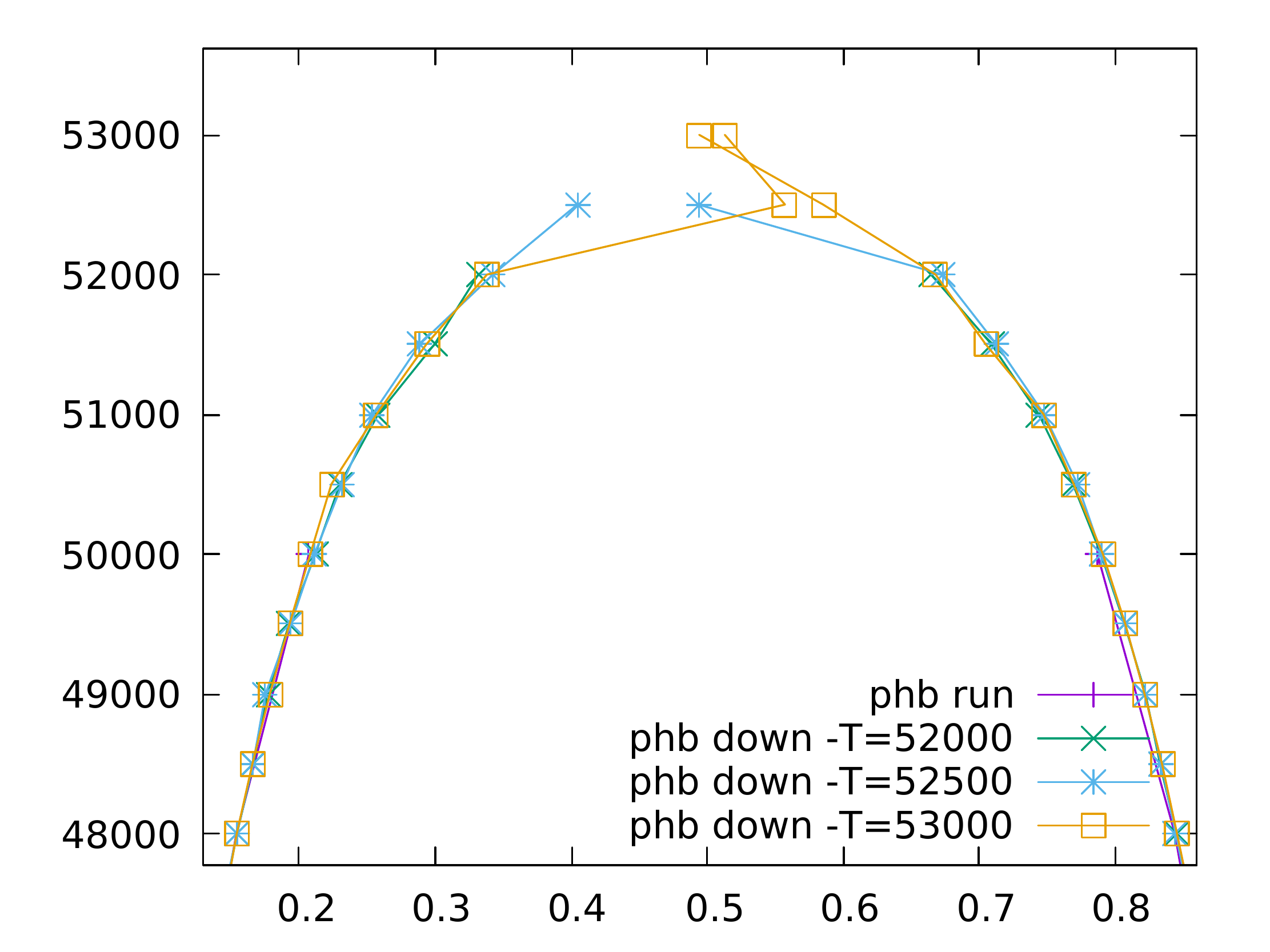}
\includegraphics[width=7cm]{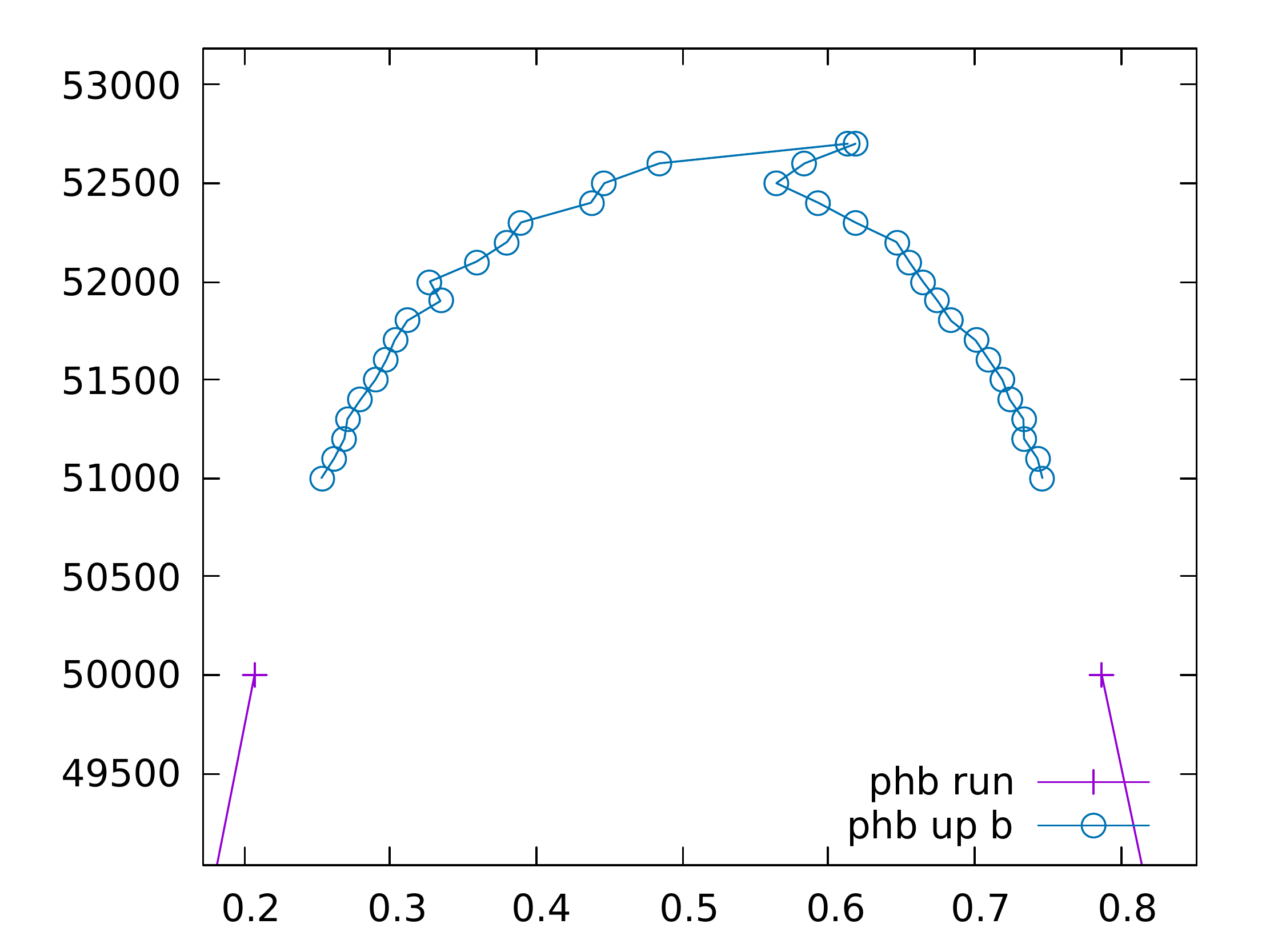}
\caption{Left: Downwards run using \phb from the center point from the
  \emc-output with three different starting temperatures
Right: Using a starting point from the downwards runs to run upwards again.
}
\label{fig:emctstat0}
\end{figure}

So we can expect a plausible endpoint of the miscibility gap to be at
$\mu=0$ somewhere around a temperature of 52000. To check
this, we can run \phb from this point \emph{downwards} to see whether we meet
our old phase separation line. To do so, I increased the size of the
cell to 50:
\begin{verbatim}
phb -gs1=0 -gs2=1 -T=52000 -mu=0 -dn -dT=500 -dx=1e-2 -er=50
    -k=8.617e-5 -ltep=5.e-3 -o=ph01-down1.out
phb -gs1=0 -gs2=1 -T=52500 -mu=0 -dn -dT=500 -dx=1e-2 -er=50
    -k=8.617e-5 -ltep=5.e-3 -o=ph01-down2.out
phb -gs1=0 -gs2=1 -T=53000 -mu=0 -dn -dT=500 -dx=1e-2 -er=50
    -k=8.617e-5 -ltep=5.e-3 -o=ph01-down3.out
\end{verbatim}
The result of these runs is shown in fig.~\ref{fig:emctstat0}.
There is still some noise at the highest point, but all three runs
actually meet with our old phase separation line. (In a downwards run,
\phb is only as good as its starting point, so it is necessary to make
sure that you reach a reasonable point.) Thus, things look quite reasonable.

We can thus use a point on this line to start upwards again (we might
also have done this directly from our first \phb-run, but I chose this
more complicated way to illustrate the possibilities), this time with
an even larger cell and a small step size for the temperature:
\begin{verbatim}
phb -gs1=0 -gs2=1 -mu=-2.94e-5 -T=51000  -dT=-100 -dx=1e-2 -er=80 
 -k=8.617e-5 -ltep=5.e-3 -o=ph01-upb.out
\end{verbatim}
With this, we get some fluctuations at the top of the miscibility gap,
see fig.~\ref{fig:emctstat0}, right,
but they are rather small. Taking all figures together, we can
estimate the temperature of the gap to about $52500\pm 1000$,
so the accuracy is about 2\%.

Finally, we can check whether results make sense at all:
According to \cite{Soisson2000}, the miscibility gap should be at
\begin{equation}
T_{\text{misc}} = \frac{0.8 \Omega_{AB}}{2 k_B} \text{ with }
\Omega_{AB}= -(z/2) (E_{AA} + E_{BB} - 2 E_{AB})\,,
\end{equation}
where $z$ is the coordination number. (According to \cite[ex. 5.10]{porter2009phase}, for a
binary system in zeroth 
order, the temperature is $\Omega/2 k_B$ regardless of lattice
structure.) For my data ($z=6$, 
$\Omega_{AB}=12$, $k=\numprint{8.617e-5}$ in units of eV), the result of this is a
temperature of about 54000, so this looks good enough.

\section{A more interesting example}
A more interesting example is one where a different phase
forms. Change the ECIs, the ground state file and the cluster
definition as follows:

\begin{verbatim}
gs_str.out
3.500000 0.000000 0.000000
0.000000 3.500000 0.000000
0.000000 0.000000 3.500000
1. 0 0
0 1. 0
0 0 1.
1.000000 1.000000 1.000000 Ni
end

3.5000 3.5000 0
3.5000 0 3.5000 
0 3.5000 3.5000 
1. 0. 0.
0. 1. 0. 
0. 0. 1
0. 0. 0. Al
0.5 0.5 0.5 Ni
end

3.500000 0.000000 0.000000
0.000000 3.500000 0.000000
0.000000 0.000000 3.500000
1. 0 0
0 1. 0
0 0 1.
1.000000 1.000000 1.000000 Al
end
\end{verbatim}
The state 0 and 2 are simply the pure elements on a simple cubic
lattice. The second is the primitive unit cell of a checkerboard
pattern (NaCl structure). 
To visualize it, you can copy this
definition into a separate file. Since the ATAT \texttt{out}-format is a bit special, I use Jesper
Kristensen's tools \cite{kristensen} to convert it into a POSCAR-format that I can open
with the \texttt{ase} tools \cite{ase}. To visualise the clusters, you
can use Jesper's script \texttt{visualize\_clusters.py} \cite{kristensen2}.


To stabilize the checkerboard pattern, we need to define more clusters
than before:
\begin{verbatim}
eci.out 
0.
0.
1
-0.2

clusters.out
1
0.000000
0

1
0.000000
1
1.000000 1.000000 1.000000

3
3.5
2
1.000000 1.000000 1.000000
1.000 1.0000 0.0000

3
7.00000
2
1.00000 1.00000 1.00000 
1.00000 1.00000 -1.00000

\end{verbatim}
Here we have changed the sign of the 2-atom cluster term so that AB as
nearest neighbours is energetically favourable.  In order to force a
third AB-phase to form, we need to make an ABA- or BAB- structure
favourable, otherwise the atoms would just dissolve freely and the
alloy has full miscibility. The final cluster in \texttt{cluster.out}
contains two atoms on the same ``color'' of the NaCl-checkerboard
pattern and the negative sign of its ECI makes a sequence AXA or BXB
favourable. So (writing 
in one dimension) a pattern of type ABABABAAAAA is better than, for
example, ABAABAAAABA because we have more identical second
neighbours if we create a separate AB-phase. 

To create these structures, you can use the \texttt{corrdump}-utility
on the \texttt{lat.in}-file; this creates a cluster-file that can be
used to define the clusters. To find all 2-atom clusters that are
smaller than a certain size limit (up to size 7 in this example), simply copy the
\texttt{lat.in}-file to a clean directory and run
\begin{verbatim}
corrdump -2=7.01
\end{verbatim}
This creates a \texttt{clusters.out}-file from which you can pick the
clusters you want.

Of course, usually you do not construct clusters and their ECIs like
this (as we will see below, this method of guessing ECIs is
dangerous), but you rather get them from another calculation (like
from ATAT's \texttt{maps} program, as we will do later). Nevertheless,
this example helps to improve understanding how \phb works and what
problems might occur.

\begin{figure}
\includegraphics[width=0.31\textwidth]{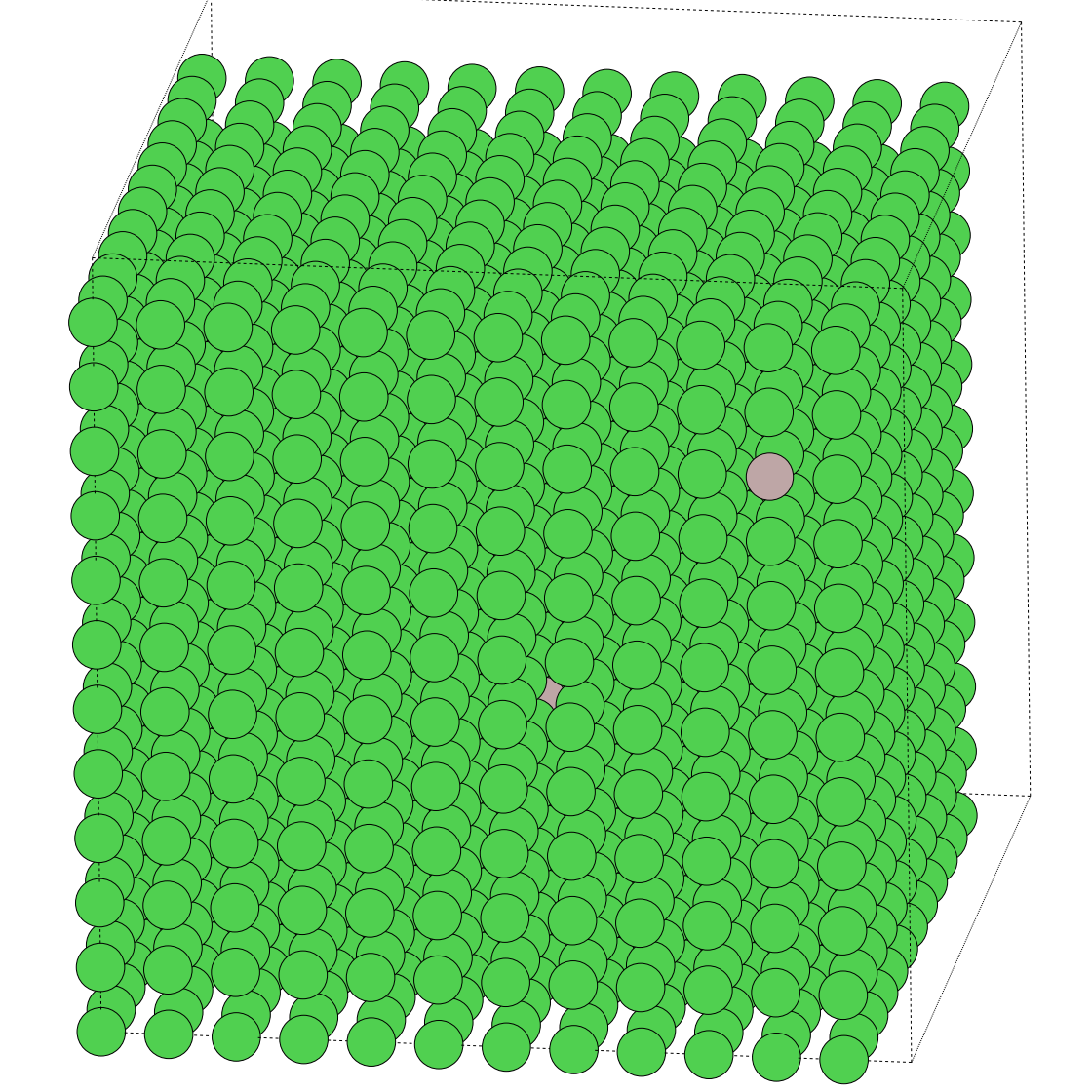}
\includegraphics[width=0.31\textwidth]{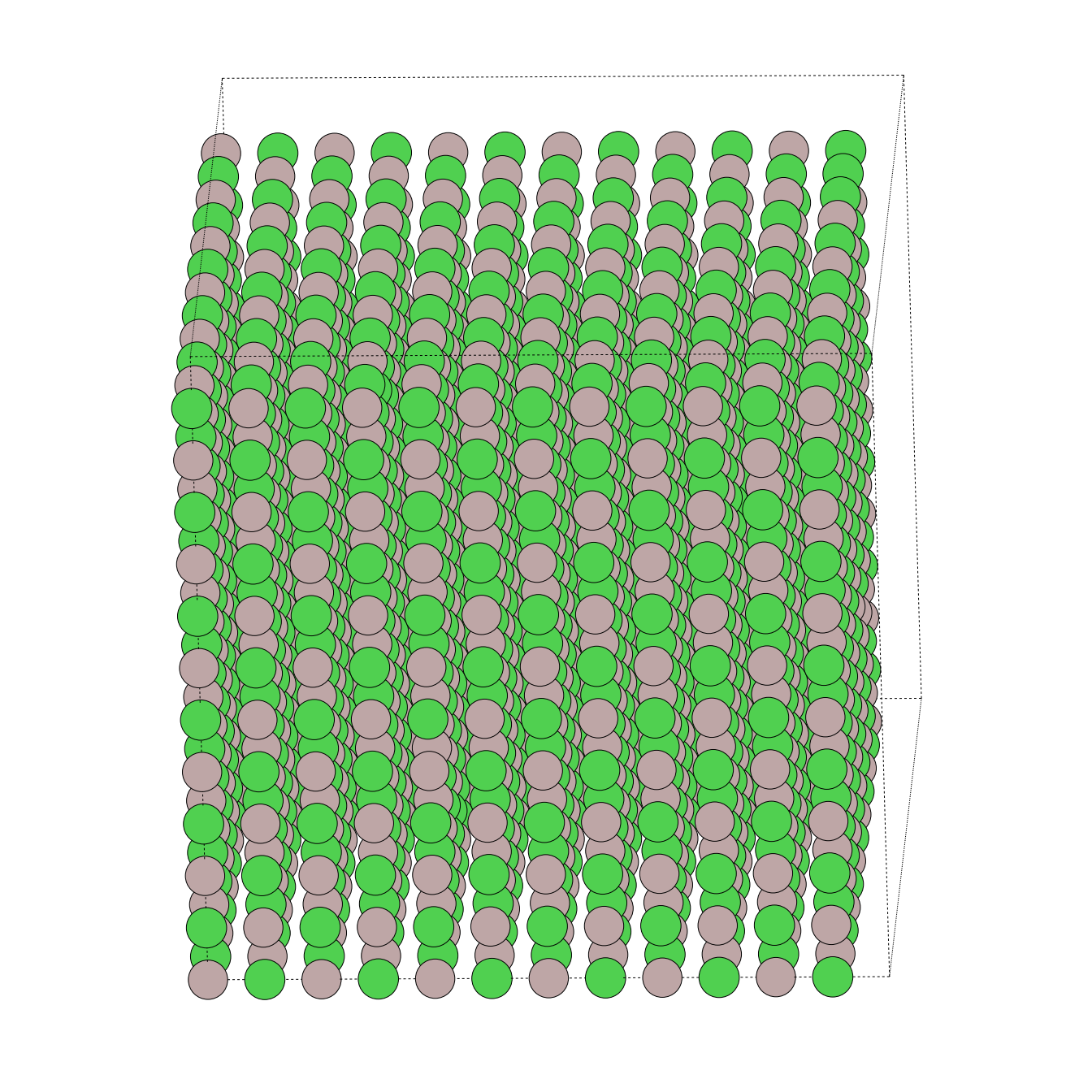}
\includegraphics[width=0.31\textwidth]{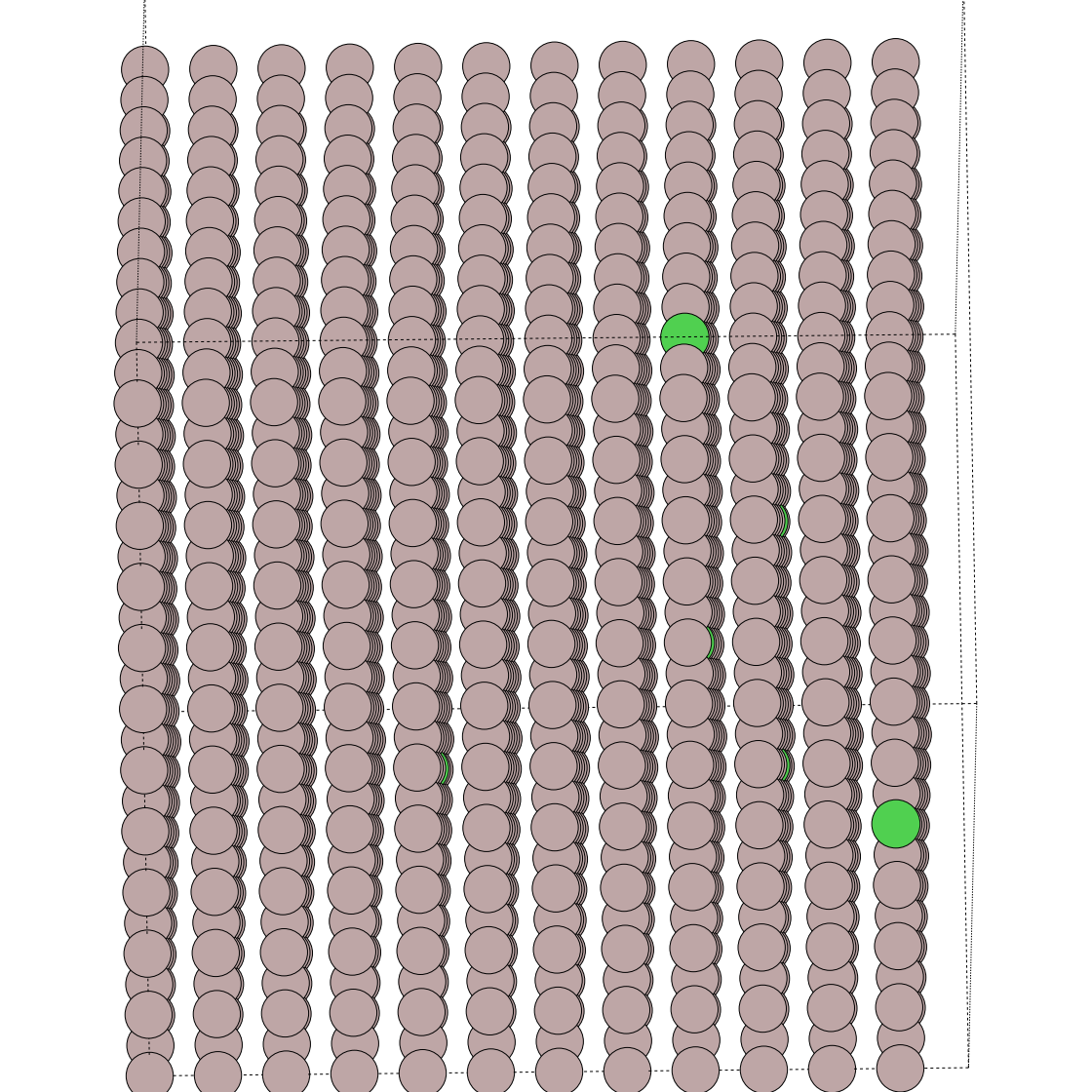}
\caption{Structures of the three phases.}
\label{fig:structs}
\end{figure}

To check that the structure definition is correct, we can create configurations
using \emc:
\begin{verbatim}
emc2 -gs=0 -mu0=0.5 -T0=30000  -keV -er=20 -dx=1-e-3
emc2 -gs=1 -mu0=1.5 -T0=30000  -keV -er=20 -dx=1-e-3
emc2 -gs=2 -mu0=2.5 -T0=30000  -keV -er=20 -dx=1-e-3
\end{verbatim}
Each run creates a configuration file \texttt{mcsnapshot.out} which we
can again plot using Jesper Kristensen's tool \cite{kristensen} to convert
out-files.  Fig.~\ref{fig:structs} shows the three structures.

You can also use \emc to understand the numbers for the chemical
potential. Simply run \emc without any MC sweeps whatsoever:
\begin{verbatim}
emc2 -gs=0 -mu0=0.5 -T0=10  -keV -er=20 -n=0 -eq=0 -g2c
Supercell size: 12 12 12
10.000000	-12.000000	2.400000	-1.000000 ......	
\end{verbatim}
and similarly for the other phases. The \texttt{-g2c}-switch tells
\emc to report quantities for the canonical ensemble; column~3 is thus
the energy per atom. This is $2.4$ (in our units): Each atom has six
nearest-neighbour bonds with energy $+1$ and six next-nearest
neighbour bonds with energy $-0.2$, resulting in an energy of $4.8$
which needs to be divided by 2 to avoid double-counting of the bonds.

The result of $-12$ for the chemical potential may seem weird. If we
are in the A-rich phase and exchange a single A-atom with B or a
B-atom with A, the total energy change is not $\pm 12$, but only
$\pm 9.6$. (6 bonds flip from $+1$ to $-1$, 6 flip from $-0.2$ to
$+0.2$.) ATAT calculates the chemical potential from the energy
difference of the provided ground states. If all ground states are
given correctly, the common-tangent construction then implies that the
chemical potential (change in energy on switching an atom in the
very-low temperature limit) can be calculated this way. (This is also
shown in Fig. 5.3 of the ATAT manual.) As we will see later (spoiler
alert), our list of ground states is incomplete, and so we have a
seeming discrepancy here.

The clusters were designed to create a phase diagram with a 3-phase
structure (A-rich, B-rich, AB ``intermetallic'') with a miscibility
gap. It should be absolutely symmetric.  Since we have two phase
boundaries, we need to do 2 \phb runs, one for the phase boundary
between phase 0 and 1 and one for the boundary between 1 and
2. Actually, I do a third run just switching 1 and 2 in the
\texttt{gs}-parameter to see what happens:
\begin{verbatim}
phb -gs1=0 -gs2=1 -dT=100 -dx=0.001 -er=50 -keV -ltep=1.e-3 -o=phase01.out 
phb -gs1=1 -gs2=2 -dT=100 -dx=0.001 -er=50 -keV -ltep=1.e-3 -o=phase12.out 
phb -gs1=2 -gs2=1 -dT=100 -dx=0.001 -er=50 -keV -ltep=1.e-3 -o=phase21.out 
\end{verbatim}
In the following, runs will be abbreviated as ``01'', ``12''
and ``21''.

Using these numbers, I would expect the 01-run and the 21-run to be
practically the same. If you compare the numbers in the output files
\texttt{phase01.out} and \texttt{phase21.out}, you see that they are
identical for temperatures up to 2700 where the LTE is valid; the only
difference between the two files is that mu is $-6$ in one case, $+6$
in the other.

At higher temperatures, the runs start to differ.
Since these are Monte-Carlo algorithms using random numbers
initialized from the system clock unless otherwise specified, two runs
with the same parameters will never be identical, but I would still
expect the runs to be very similar. This, however, does not happen.
The 01-run proceeds up to
a temperature of 63400; the largest number of MC
iterations is 32000. (The exact number may vary depending on the
random numbers in your MC sequence.) 
The 21-run (which should be identical) becomes very slow, at a temperature of 13200, we get 
\begin{verbatim}
Phase 1 n_equil= 64000 n_avg= 896000
Phase 2 n_equil= 0 n_avg= 2000
\end{verbatim}

The 12-run
shows another problematic behaviour: At a temperature of 13200, we get
the following output:
\begin{verbatim}
13200	5.96865	0.450491	0.746105	-3.34662	-3.35492	
Phase 1 n_equil= 0 n_avg= 4000
Phase 2 n_equil= 0 n_avg= 3000
0	1
Looking for phase transition...
mu           x
5.992864	0.768726
6.016864	0.782460
...
45.976864	1.000000
46.000864	1.000000
\end{verbatim}
The program stalls and never proceeds to higher temperatures.
This may happen when program ``wanders outside of the region
of metastability'' \cite[p. 15]{Walle2002} and tries to find the correct mu-value.

\begin{figure}
\includegraphics[width=6cm]{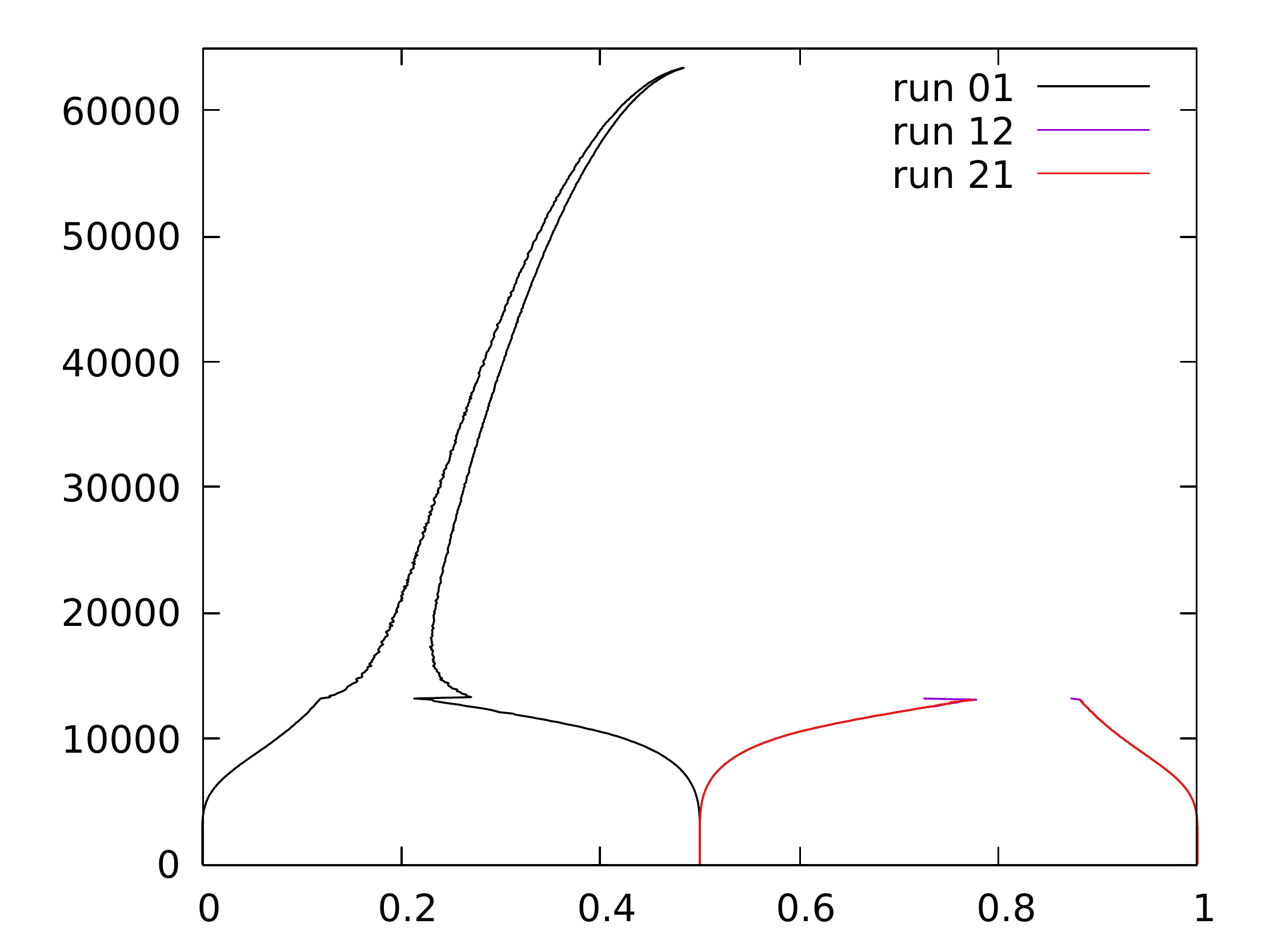}
\includegraphics[width=6cm]{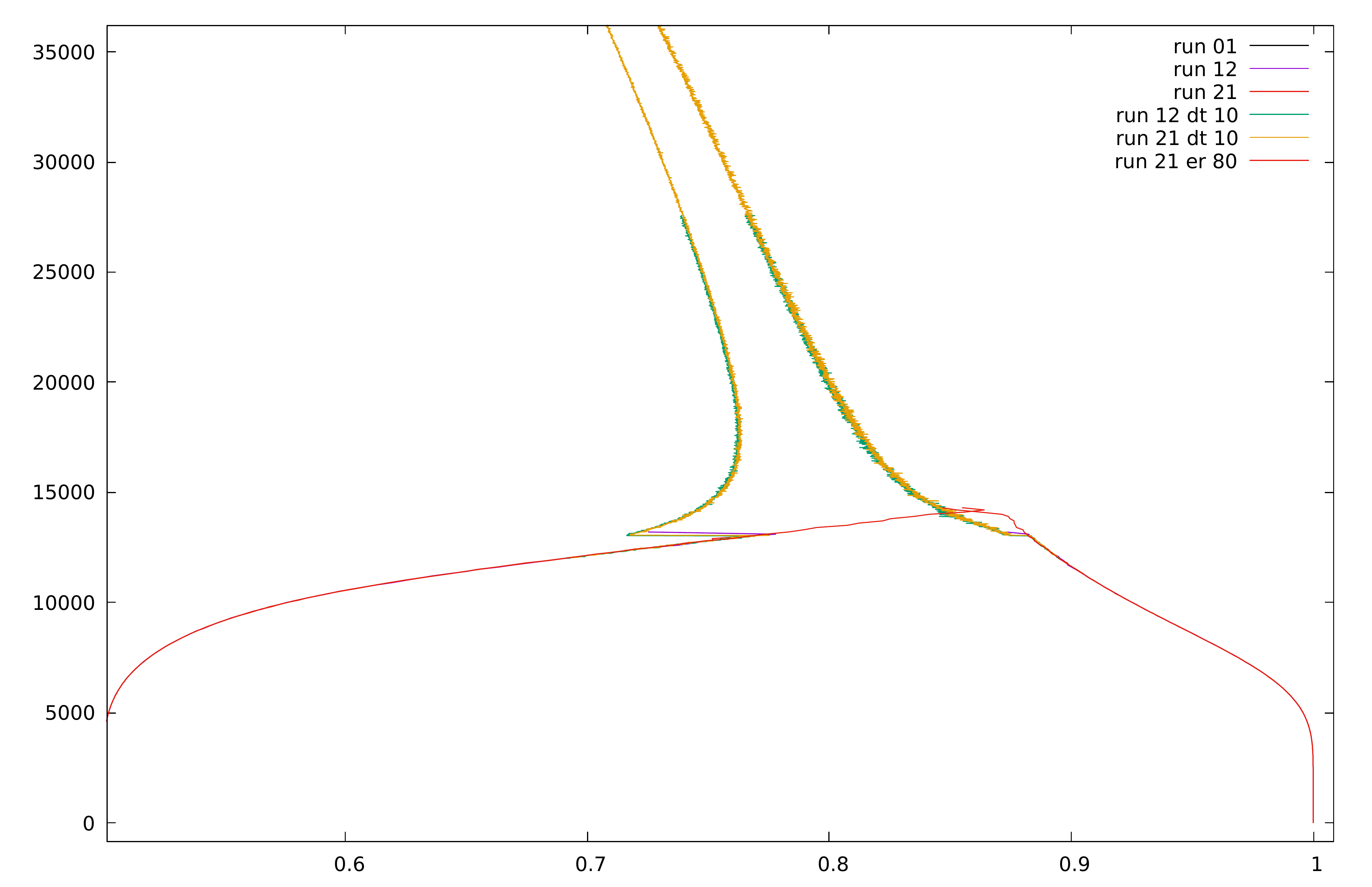}
\caption{Left: Calculation of the phase diagram with \phb for the 3-phase
  alloy. Note that the 12- and 21-runs are stalling whereas the 01-run
  calculates everything up to the miscibility gap.
Right: Zoom of the calculation re-done with smaller temperature steps
and with larger radius.
}
\label{fig:newstruct1}
\end{figure}

The resulting phase diagram is shown in fig.~\ref{fig:newstruct1}, left.
Only one of the runs  calculates the full 2-phase region
up to the high-temperature state. 
It is also interesting that there are strange 
``spikes'' at concentrations of about 0.25 and 0.75 that are possibly
due to \phb not correctly tracking the phase boundary. (More on these below.) At the same
temperature, there are smaller discontinuities at concentrations of about 0.12 and 0.88. 

To study this further and get a better result in the right-hand part
of the diagram, we can re-run \phb with smaller step size for \texttt{dT}. To do
so, use values for $T$ and $\mu$ from the out-file of the 12-run (at
temp 12000, the value for the 21-run differs only in the last digit) 
\begin{verbatim}
12000  5.97369   0.380374   0.790879 ....
\end{verbatim}
and start \phb from this point (I could of course start from the LTE,
but then the calculation would take much longer, even so, the
following runs need several hours):
\begin{verbatim}
phb -gs1=1 -gs2=2 -T=12000 -mu=5.97369  -dT=10 -dx=0.001 -er=50 -keV 
 -ltep=1.e-3 -o=phase12-dt10.out
phb -gs1=2 -gs2=1 -T=12000 -mu=5.97369 -dT=10 -dx=0.001 -er=50 -keV 
 -ltep=1.e-3 -o=phase21-dt10.out 
\end{verbatim}
I also run it again with a larger cell (\texttt{er}):
\begin{verbatim}
phb -gs1=0 -gs2=1 -dT=2000 -dx=1e-2 -er=80 -k=8.617e-5 -ltep=5.e-3
-o=ph01-er80.out
\end{verbatim}
The results are shown in \ref{fig:newstruct1}, right. The smaller
temperature step reproduces the behaviour from the other side of the
concentration region, including the strange spikes in the curve. They
are also a bit strange since the line is not smooth but seems to
oscillate. The
run with larger radius seems to close the phase diagram, but although
the two red lines cross, \phb does not finish the calculation.

So things look really strange here. To find out what is going wrong,
we can look at the most critical aspect of the behaviour of the
simulation: The spike at concentrations 0.75/0.25. Perhaps something
interesting is happening at these concentrations?

To find out, we can use \emc. Since ATAT always gets chemical
potentials as input variables, not concentrations, it is difficult to
fix a certain concentration value.\footnote{You can specifiy a
  concentration with \texttt{-x} in \emc, but this only affects the
  initialization.} 
We could of course change the
source code (it would probably not be too difficult to use the
existing routines to write a program that does a canonical MC simulation), but we can
also simply try to guess a good value for the chemical
potential. Since $\mu=2.5$ (input variable $\mu$) perfectly stabilizes pure B (concentration 1.0) and 
$\mu=1.5$ stabilizes the AB-phase at concentration 0.5, a value around
2.0 should be fine. We run \emc with a small cell at several $\mu$-values in
this region at a very low temperature. We use a large number of
equilibration steps and only one actual measurement step so that the
output file describes the final state. We start at a disordered state
and ignore the phase structure (\texttt{tstat=0}) so that \emc can try
to simply find the lowest-energy structure. So run several commands of
this type:
\begin{verbatim}
emc2 -gs=-1 -mu0=1.96  -T0=1000  -keV -er=10 -n=1 -eq=500000 -tstat=0
\end{verbatim}
with different values of \texttt{-mu0} and check the output file. And
indeed, at a value of 1.96, we find a concentration (in ATAT units) of
0.5 for the final state (physical concentration of 0.75). 
At \texttt{-mu0} of 1.97, we find another structure which has a
concentration of 0.75 (0.875 in physical units). 
Note that these runs are sensitive to random
number initialization, so you may find slightly different values for \texttt{-mu0}
to stabilize the structures we are looking at. Depending on the
numbers you might also find other stable structures.\footnote{This
  also explains why we got some seemingly inconsistent result for the
  calculation of the chemical potential at the beginning of this section.}

\begin{figure}
\includegraphics[width=6cm]{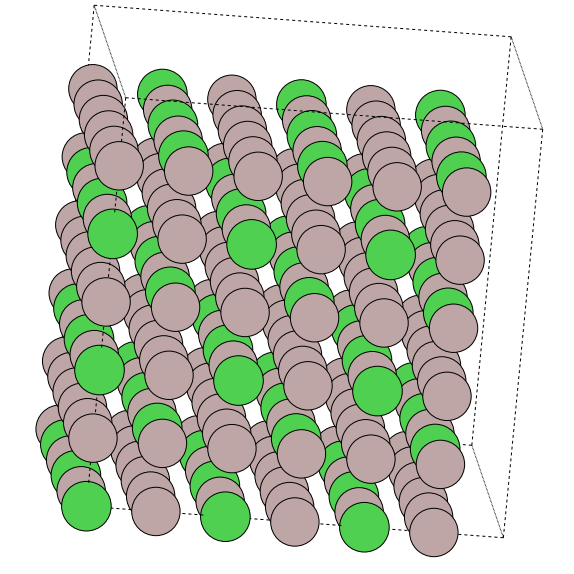}
\includegraphics[width=6cm]{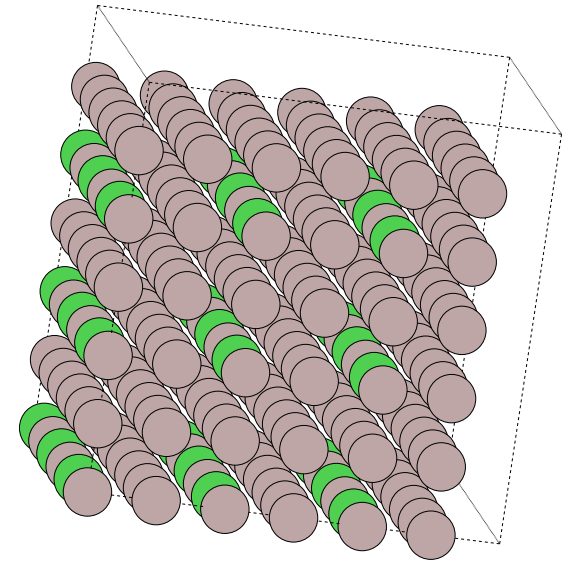}
\caption{Stable structures with concentrations of 0.75 and 0.875.}
\label{fig:otherStruct}
\end{figure}

To understand these structures, we can plot the end states using the
\texttt{mcsnapshot.out} files, see fig.~\ref{fig:otherStruct}. Looking
at these structures, we see that each A-atom is surrounded only by
B-atoms as nearest neighbours (the optimal structure for the
nearest-neighbour cluster) and that each atom also has another
atom of the same species as next-nearest neighbour along each axis. So these structures
are also stable with our cluster definitions. 

Thus, the problem in this example lies not with \phb, but
with my incorrect definition of stable structures; there are stable
structures that were not included in my \texttt{gs\_str.out}-file. The
weird behaviour of \phb was thus due to structures interfering that
were not included in the definition.

\section{Calculating a phase diagram from first principles}
\subsection{The Ni-Al phase diagram: First attempt}
It seems that guessing cluster coefficients is not so easy~-- so
let's do a density functional theory calculation to get a meaningful
cluster expansion. This is done using \maps. Since I'm working a lot
with nickel alloys, we can look at the NiAl-system.

The basic idea is explained nicely in the ATAT manual:
\maps is the control program that creates new configurations to
calculate, \pollmach is the program that manages the calls to the
first principles code (VASP in my case) and this calls \rsv. \rsv
calls another program called \ezvasp that performs
the actual runs. 

Before you start, you need a \texttt{\textasciitilde/.ezvasp.rc}-file in your home
directory that tells \ezvasp where the potentials are and how VASP is
to be called. My file looks like this:
\begin{verbatim}
#!/bin/csh
#enter name of vasp executable here
set VASPCMD="mpirun -n 12 vasp"
#enter the directories containing the pseudopotentials here
set POTPAWLDA="/opt/vasp/potentials/potpaw_lda/"
set POTPAWGGA="/opt/vasp/potentials/potpaw_gga/"
set POTPAWPBE="/opt/vasp/potentials/potpaw_pbe.5.4/"
set POTLDA=$POTPAWLDA
set POTGGA=$POTPAWGGA
\end{verbatim}
As you can see, I want to call vasp with \texttt{mpirun} and run on 12
of my 16 cores. I do run in serial mode since I am doing all this on a
simple workstation, not on a massively parallel machine. The VASP
potentials are situated in the \texttt{/opt}-directory. Note that
\texttt{ezvasp -h} gives you lots of informations on how to do the
VASP runs, but note also that you do not directly interface \ezvasp
but only do it with \rsv. For example, \ezvasp tells you that you can
write magnetic moments in the POSCAR section of the file \ezvasp gets,
but as far as I can see you cannot do this if you call it from
\rsv via \pollmach.

To do the calculation, you need a \texttt{lat.in}-file as before and a
wrapper file. Here is my \texttt{vasp.wrap}:
\begin{verbatim}
[INCAR]
PREC = high
ISMEAR = 1
SIGMA = 0.07
NSW=41
IBRION = 2
ISIF = 3
KPPRA = 1000
USEPOT = PAWPBE
NPAR=4
ISPIN=2
SUBATOM = s/Ni$/Ni_pv/g
EDIFF = 1e-8
EDIFFG = 1e-5
DOSTATIC
\end{verbatim}
I have taken most of the parameters from the ATAT example. Note that I
use the \texttt{SUBATOM}-command that is documented in the
\ezvasp-help. This allows to replace an element name with the name of
the potential -- for Ni, it is good to include the $p$-electrons and
use the \texttt{Ni\_pv}-potential. Note also that I use tight values
for \texttt{EDIFF} and  \texttt{EDIFFG}. The smaller value for
\texttt{SIGMA} is mainly for consistency with previous calculations I
have made with nickel.

Actually, I should also have entered a sensible \texttt{ENCUT},
otherwise, the pure Al run in directory \texttt{1/} has a
different cutoff from the other simulations. Checking afterwards
showed that the energy of the Al run was only affected slightly by
this. Still, I strongly recommend to always set ENCUT (which is a good
practice for VASP in general).

Nickel is ferromagnetic, so I also include \texttt{ISPIN=2}. You
cannot specify a \texttt{MAGMOM} in the wrapper-file because this gets
directly copied to the later \texttt{INCAR}-file and as soon as the
first 2-particle calculation is done, VASP complains about the
\texttt{MAGMOM-line}. However, for nickel, this should not matter much because
the default initialization of~1 for \texttt{MAGMOM} is fine. (If you
do not include the spin polarization, ATAT finds a stable
Ni${}_7$Al-phase as described briefly in \cite{Woodward2014}.)


For completeness, here is the simple \texttt{lat.in} with a reasonable
value for the lattice constant:
\begin{verbatim}
3.52 3.52 3.52 90 90 90
0   0.5 0.5
0.5 0   0.5
0.5 0.5 0
0.000000 0.000000 0.000000 Ni,Al
\end{verbatim}

To start everything, do (after making sure that your VASP settings are
good enough to give precise results)
\begin{verbatim}
maps -d &
touch ready
pollmach runstruct_vasp &
\end{verbatim}
and then \maps runs and creates configurations. The \maps output is
again explained well in the manual. Note that you can find the ground
states of the system not only in \texttt{gs\_str.out} in the same
format as before, but also in \texttt{gs.out} where the structure
number (number of the subdirectory) is given. (So if you want to
visualise the ground states, you can use the \texttt{POSCAR}-files in
the directories listed in \texttt{gs.out}.) The file also
lists the energy the structure actually has and the energy you get
from the cluster expansion. 

If you get errors like (in your system language)
\begin{verbatim}
cp: cannot stat 'OSZICAR': No such file or directory
cp: cannot stat 'OUTCAR': No such file or directory
\end{verbatim}
something is wrong with the VASP call; \rsv does not directly
check whether VASP has run so the code only notices this when it tries
to copy some files. Check your VASP installation or the interface
between \maps and VASP.

During the run, \maps reports on the performed calculations; don't
worry if you see lines like
\begin{verbatim}
Finding best cluster expansion...
1 1 1 1 3.40282e+38
\end{verbatim}
the \texttt{FLOATMAX} is no problem.

\begin{figure}
\includegraphics[width=0.48\textwidth]{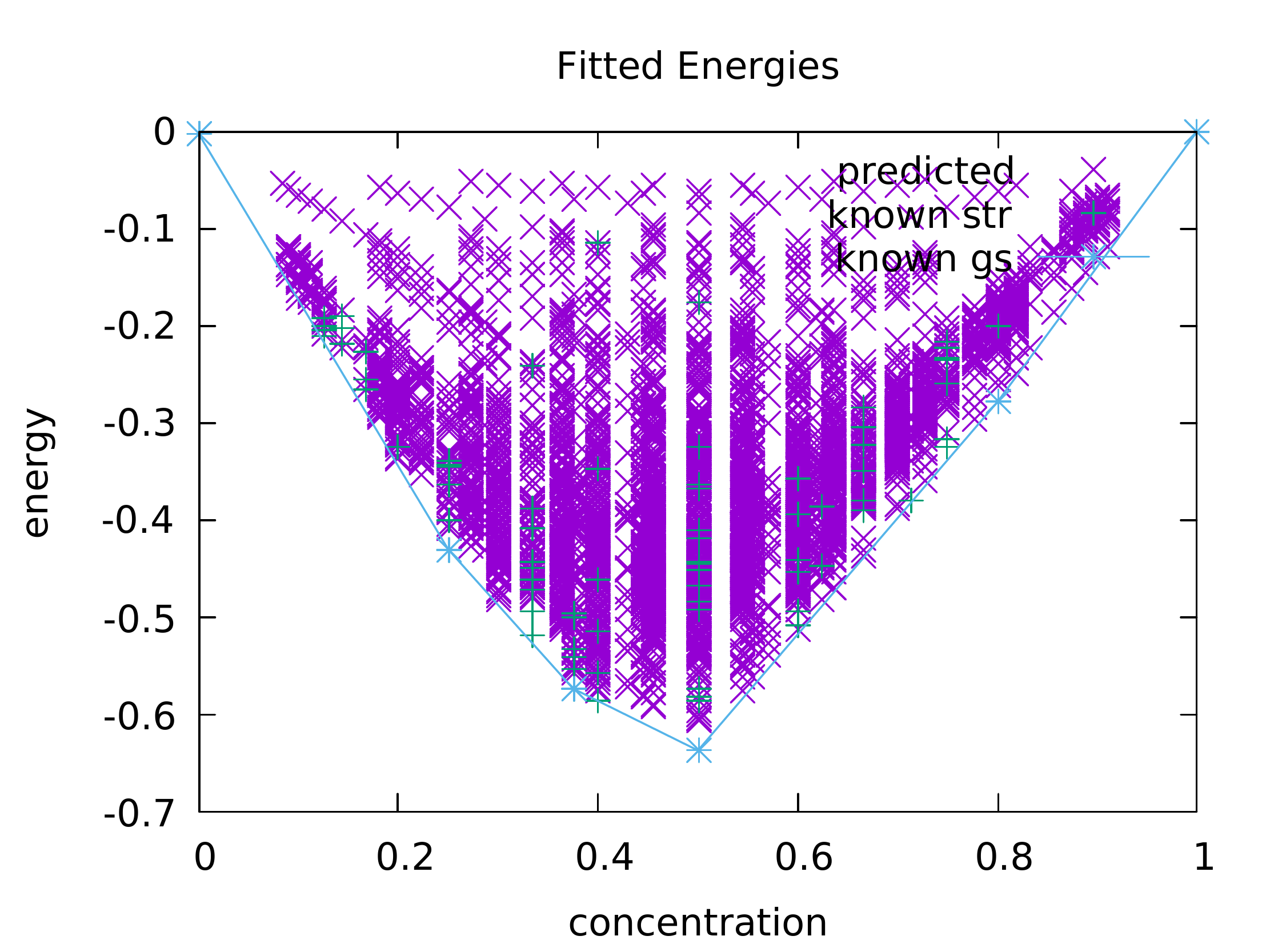}
\caption{ATAT calculation result plotted with \texttt{mapsrep} for the
  NiAl system.}
\label{fig:mapsrep}
\end{figure}

After about two days of running (on my machine, using 12 Xeon
processors), ATAT finds four phases (beyond the pure Ni and Al phase):
\ch{Ni_3Al}, \ch{Ni_5Al_3}, \ch{NiAl}, \ch{Al_4Ni}. Comparing this
to the Ni-Al phase diagram, it seems that we get the nickel-side
correct, the Al-side should have an \ch{Ni_2Al_3}- and an
\ch{Al_3Ni}-phase. A look at the energies with \texttt{mapsrep}
(figure~\ref{fig:mapsrep}) shows 
that \ch{Al_4Ni} is only barely stable. (The data point lies almost
exactly on the common tangent between Al and NiAl.)
We stop the simulation with
\texttt{touch stoppoll}. Wait until the current VASP run is finished
and \maps has updated the results, then
\texttt{touch stop}. 

To analyze the results,  we can also use the program
\texttt{checkrelax}. This calculates how distorted the relaxed cells
of the different runs actually are (it measures the strain in the
cells). Do
\begin{verbatim}
checkrelax > checkrelaxresult
\end{verbatim}
and look at the output file \texttt{checkrelaxresult}. In our case, some of the cells are very
strongly distorted, the file ends with:
\begin{verbatim}
0.2024 54
0.2069 3
0.2115 46
0.2186 8
0.2456 48
0.2535 25
\end{verbatim}
If strains are very large, the result is not really useful for the
cluster expansion, because in doing the MC runs later on, we assume
that things happen on our prescribed lattice. Expansion coefficients
calculated from a strongly distorted cell may thus be incorrect. 

In principle, we can exclude all those configurations. (See
section~\ref{secDistort} below on how to do that.)
However, in our case there is a problem: Configuration~3 is a ground
state as you can see by looking into \texttt{gs.out} which lists all
the ground states. If you look into the corresponding \texttt{POSCAR}
in directory \texttt{3}, you see that this is the
\ch{NiAl}-phase, but has an orthorhombic cell instead of the
correct cubic CsCl-structure. The reason for this problem is simple:
The CsCl-structure is a bcc-structure, not an fcc-structure, and thus
cannot be found with the current setting of \texttt{lat.in}. ATAT
creates an initial fcc-lattice that is occupied by alternating Ni and
Al atoms and this then gets distorted to a bcc lattice. (This is
reminiscent of the Bain model of the martensitic transformation in
iron \cite{rosler2007mechanical}.)

In any case, if you do this exclusion and calculate a phase diagram (I
tried), you will find that your phase structure looks simply awful and
that the calculation with \phb frequently stalls.  If you actually
need the full Ni-Al phase diagram, you could do a \maps
calculation with a bcc lattice as well; you would then have to do the
phase diagram calculations with several different phases.

However, for this tutorial we choose a simpler way.

\subsection{Calculating the Ni-rich part of the phase diagram }
In \maps, you can specify the concentration range you are interested
in via the command-line options \texttt{-c0=a -c1=b}.
Let us restrict ourselves to the calculation of the phase boundary
between Ni and \ch{Ni_3Al}  to avoid trouble
with the bcc phase. Create a new directory, set up
\texttt{lat.in} and \texttt{vasp.wrap} as before and then call maps as 
\begin{verbatim}
maps -d -c0=0 -c1=0.25 &
touch ready
pollmach runstruct_vasp &
\end{verbatim}
(Instead of doing all the calculations again, it would probably be easier to
copy the old result directories for those structures that are in the
right part of the phase diagram, but I prefer to do a clean start.) 
After a weekend of calculations, we can stop the calculation with
\texttt{touch stoppoll} and \texttt{touch stop}. 


From this calculation, we get six 2-particle clusters, two 3-particle
clusters and one 4-particle cluster (this is different from
\cite{Woodward2014} where some ``special quasi-random
structures''\footnote{Special quasi-random structures are supercells
  that are designed to correctly reproduce the cluster
  structure. You can generate them with the \texttt{mcsqs}-program
  that is distributed with ATAT. You provide the maximum size of
  clusters of 2, 3 or more particles and the cell size you would like
  and a Monte Carlo procedure is used to arrange atoms in the way that
optimally represents the cluster coefficients.} are
used as well to optimize the cluster calculation.). You get this
information most easily from the
\texttt{clusinfo.out}-file:\footnote{Be careful with this file; it is
  not created by \maps but only by \texttt{mapsrep}.}
\begin{verbatim}
2 2.489016 6 0.096360
2 3.520000 3 0.002918
2 4.311102 12 0.003826
2 4.978032 6 -0.003782
2 5.565609 12 -0.001564
2 6.096819 4 0.004159
3 2.489016 8 -0.005556
3 3.520000 12 0.002930
4 2.489016 2 -0.025176
\end{verbatim}
First number is the number of particles, second the size, third the
number of equivalent configurations, and fourth the ECI. Note that the
file \texttt{eci.out} contains two more entries for the zero- and
one-particle ``cluster''.

\begin{figure}
\includegraphics[width=0.5\textwidth]{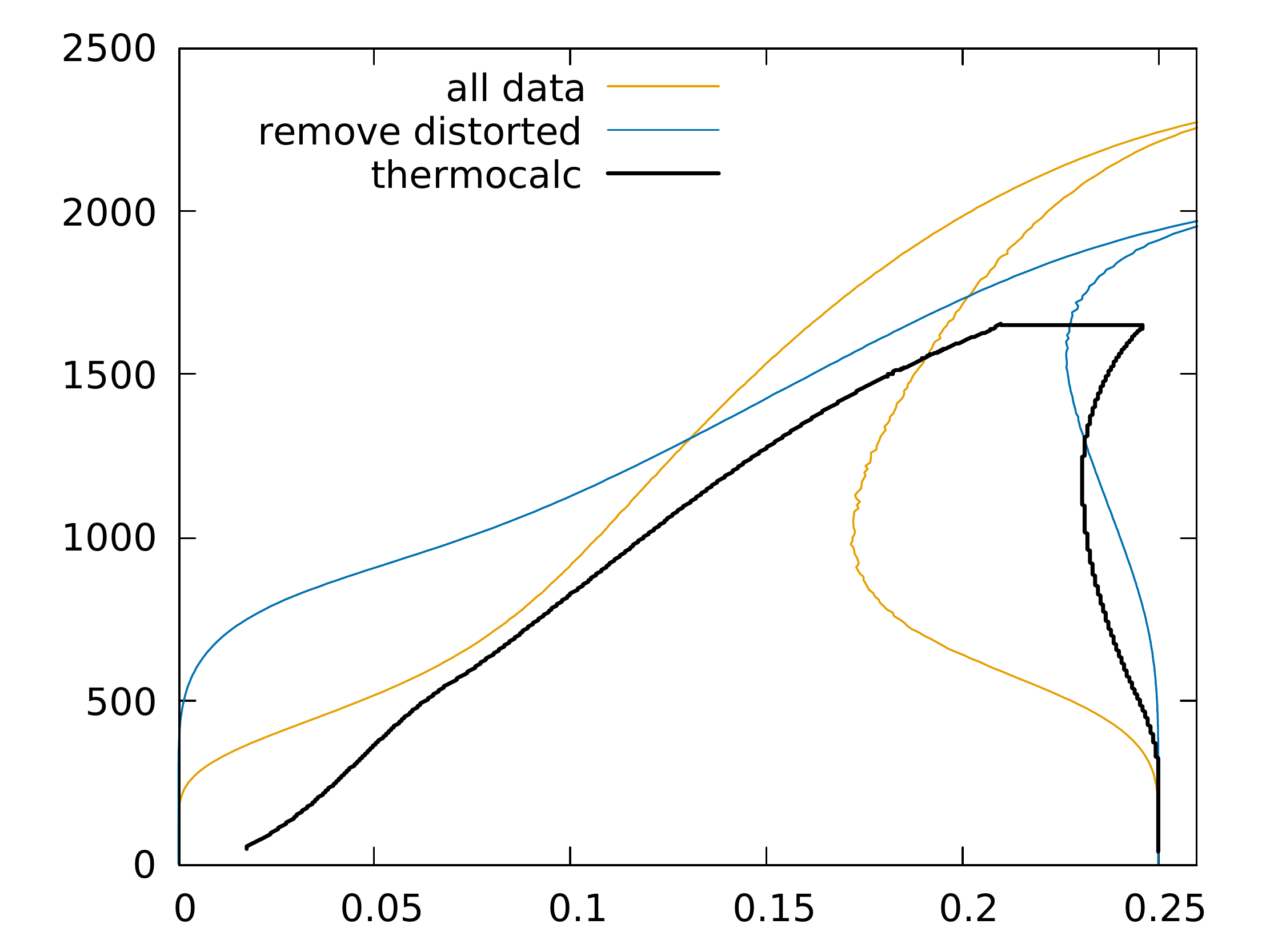}
\caption{Calculated Ni-rich part of the Ni-Al phase
  diagram. The orange line marked ``all data'' describes the first run in
  this section, the blue line is discussed in
  section~\ref{secDistort}. The thick black line is the calculated phase diagram using the TTNi8 database of
  Thermocalc \cite{thermocalc}.}
\label{fig:nialPhase1}
\end{figure}

Copy \texttt{lat.in} and all the out-files to a new directory (you can run in the same
directory without trouble, but I would recommend to keep files
separated) and run \phb as usual:
\begin{verbatim}
phb -gs1=0 -gs2=1 -dT=10 -dx=0.001 -er=50 -keV -ltep=1.e-3
 -o=phase01.out
\end{verbatim}
This may take a while, but in the end you get a rather nice phase
diagram of the Ni-rich part of the phase diagram, see
Fig.~\ref{fig:nialPhase1} (the yellow curve).
Of course the high-temperature behaviour is wrong (melting is not
included), but overall the diagram does at least look realistic. It
differs from a full calculation done with thermodynamic software (I
used Thermocalc \cite{thermocalc} in the figure), but it
looks similar to the diagram shown in \cite{Woodward2014}.

\subsection{Remove distorted configurations}\label{secDistort}
We did not check whether there are distorted configurations that may
affect the results of the computation.  I recommend doing this by
copying your whole maps-directory (where all the DFT calculations
reside) of the run to a new name so that you can easily compare old
and new runs. Create a \texttt{checkrelaxresult}- file with
\texttt{checkrelax > checkrelaxresult} . It contains 5 configurations
that have a value larger than 0.1:
\begin{verbatim}
0.1062 609
0.1090 51
0.1119 13
0.1666 24
0.2220 608
\end{verbatim}

To remove these, we create a file called \texttt{error} in the
directory of the corresponding run. For example, \texttt{touch
  609/error} removes the result from the most strongly distorted cell
from the calculation.  We can exclude all those directories where
\texttt{checkrelax} gives a value larger than 0.1 (the value
recommended in the man page) by simply editing the output from
checkrelax, replacing the first number with the string \texttt{touch}
and appending an \texttt{/error} at the end. I do this with an
\texttt{emacs} keyboard macro, but you can also use \texttt{awk},
\texttt{sed} or whatever your favorite tool for these things is, if
the number is annoyingly large. (However, if it is, it is probable
that something is wrong with your lattice definition as in the case of
the NiAl-phase above.) If you do this, don't forget to re-run \maps
(and \texttt{touch stop} after it finishes) after you touched the
error files so that the new ECIs are actually
calculated.\footnote{When you re-run \maps, the logfile may state that
it only looked at ground states with 0 atoms: \texttt{The internal
  database of structures extends at least up to 0 atoms/unit cell}.
The cluster calculation is nevertheless done correctly. If you want to
ensure that everything looks o.k., you can use the
\texttt{-gs=}-parameter of \maps to specify the number of atoms in the
ground state to be looked at.}

We can then do the same phase diagram calculation as before. The
result is also shown in fig.~\ref{fig:nialPhase1}. The
right-hand phase separation line is much closer to the Thermocalc
prediction; on the left-hand side, the low-temperature region is a bit
worse but the high-temperature behaviour looks quite good.
Probably, we could fine-tune the calculation here by adding
some configurations from the calculation, using SQS.

\subsection{Phonon and electron influence}

In this step, we follow the \texttt{steps.txt}-file that is in ATAT's
\texttt{tutorial} directory to calculate temperature-dependent ECIs
that contain the influence of phonons and of electronic degrees of
freedom.  We use the version from the previous section with removed
distorted configurations.


Create a \texttt{force.wrap}-file, for example:
\begin{verbatim}
[INCAR]
PREC = high
ISMEAR = 1
SIGMA = 0.07
NSW=0
KPPRA = 1000
USEPOT = PAWPBE
NPAR=4
ISPIN=2
SUBATOM = s/Ni$/Ni_pv/g
EDIFF = 1e-8
EDIFFG = 1e-5
\end{verbatim}
If you forget to create this file, \texttt{pollmach} will warn you:
\texttt{You need a force.wrap file in one of the directories ., .. ,
  ../.. , etc.}. 

\begin{figure}
\includegraphics[width=7cm]{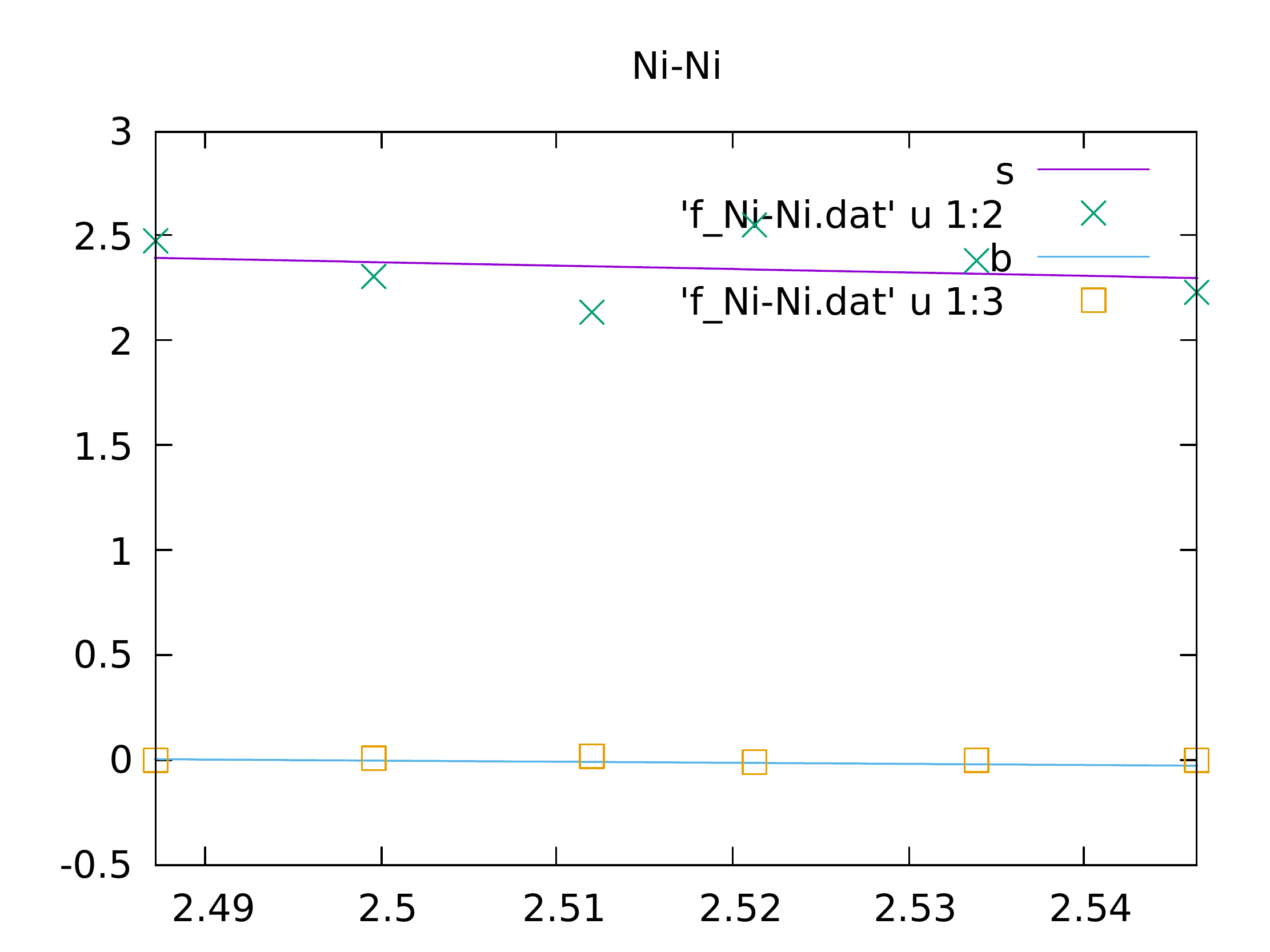} 
\includegraphics[width=7cm]{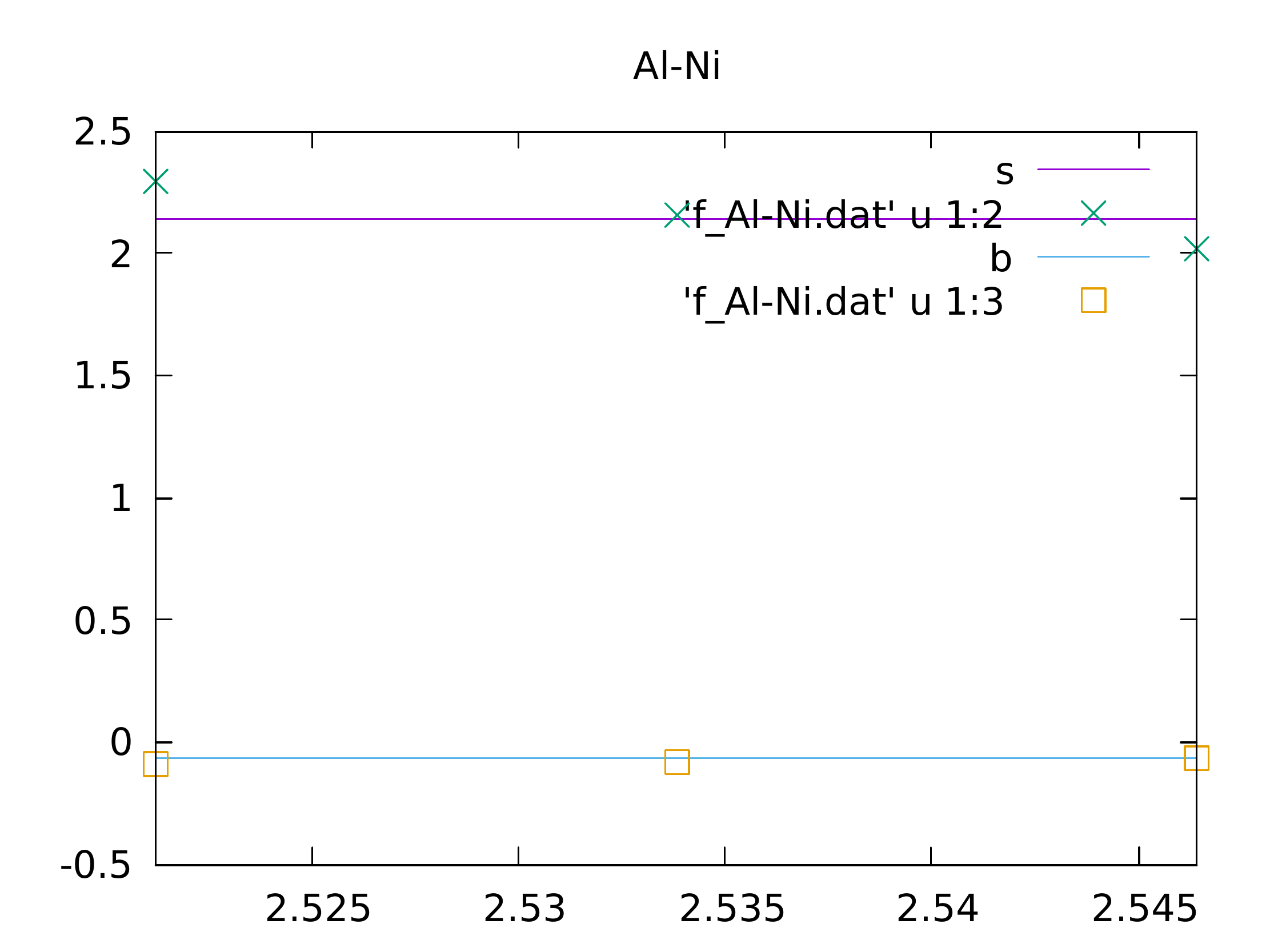} 
\caption{Calculated force-distance relation (linear fit) from
  \texttt{fitsvsl}. 
}
\label{fig:svsl}
\end{figure}

To find out the numbers of the ground states we need to calculate the
phonons for, look into \texttt{gs.out}. There are only two, one for
pure Ni, one for \ch{Ni_3Al}. For
\texttt{fitsvsl}, I evaluate 3 different volumes instead of only~2 and
I set the distance of images to~8
because the manual states
``Typically, \texttt{-er} should be 3 or 4 times the nearest neighbor distance''.
Possibly, a larger value might work better; to get quantitatively
reliable results, it would probably be wise to try different
values and check how they influence the results.
\begin{verbatim}
echo 0 27 > strname.in
fitsvsl  -ns=3 -er=8 
pollmach -e runstruct_vasp -w force.wrap &
fitsvsl -f
gnuplot fitsvsl.gnu
\end{verbatim}
The \gnuplot-command plots the force-distance relation, it looks not
perfect in this case (figure~\ref{fig:svsl}), but reasonable.

We continue as in the tutorial
\begin{verbatim}
echo 2000 21 > Trange.in
foreachfile -e str_relax.out pwd \; svsl -d   
clusterexpand -e svib_ht
clusterexpand -e fvib
\end{verbatim}
This creates the ECIs for the vibrational DOFs in the desired
temperature range. With
\begin{verbatim}
mkteci fvib.eci
\end{verbatim}
we can create a teci.out-file that contains temp-dependent ECIs.

Now also add electronic dofs:
\begin{verbatim}
foreachfile -e str_relax.out pwd \; felec -d
clusterexpand -e felec
mkteci fvib.eci felec.eci
\end{verbatim}
This first uses the \texttt{foreachfile}-utility to do a \texttt{felec} calculation
in every directory that contains \texttt{str\_relax.out}. Note that the option
\texttt{-e} ignores directories that contain an error file; if this is
not included and error file are present, the command may stall.

For low temperatures, the temperature-dependent coefficients are
similar to those in \texttt{eci.out} as they should be:

\begin{tabular}{ll}
\texttt{teci.out}&\texttt{eci.out}\\
-0.392042     &  -0.425542  \\
-0.057159     &  -0.044139    \\
0.116381      &  0.118159     \\
-0.004468     &  -0.004468    \\
0.003976      &  0.003991     \\
-0.00671667   &  -0.006715    \\
-0.00670783   &  -0.006762    \\
-0.0044705    &  -0.004538    \\
-0.00478617   &  -0.004812    \\
0.0134192     &  0.013498     \\
\end{tabular}

The first value differs considerably, I assume that  this is due to
the zero-point energy.

\begin{figure}
\includegraphics[width=7cm]{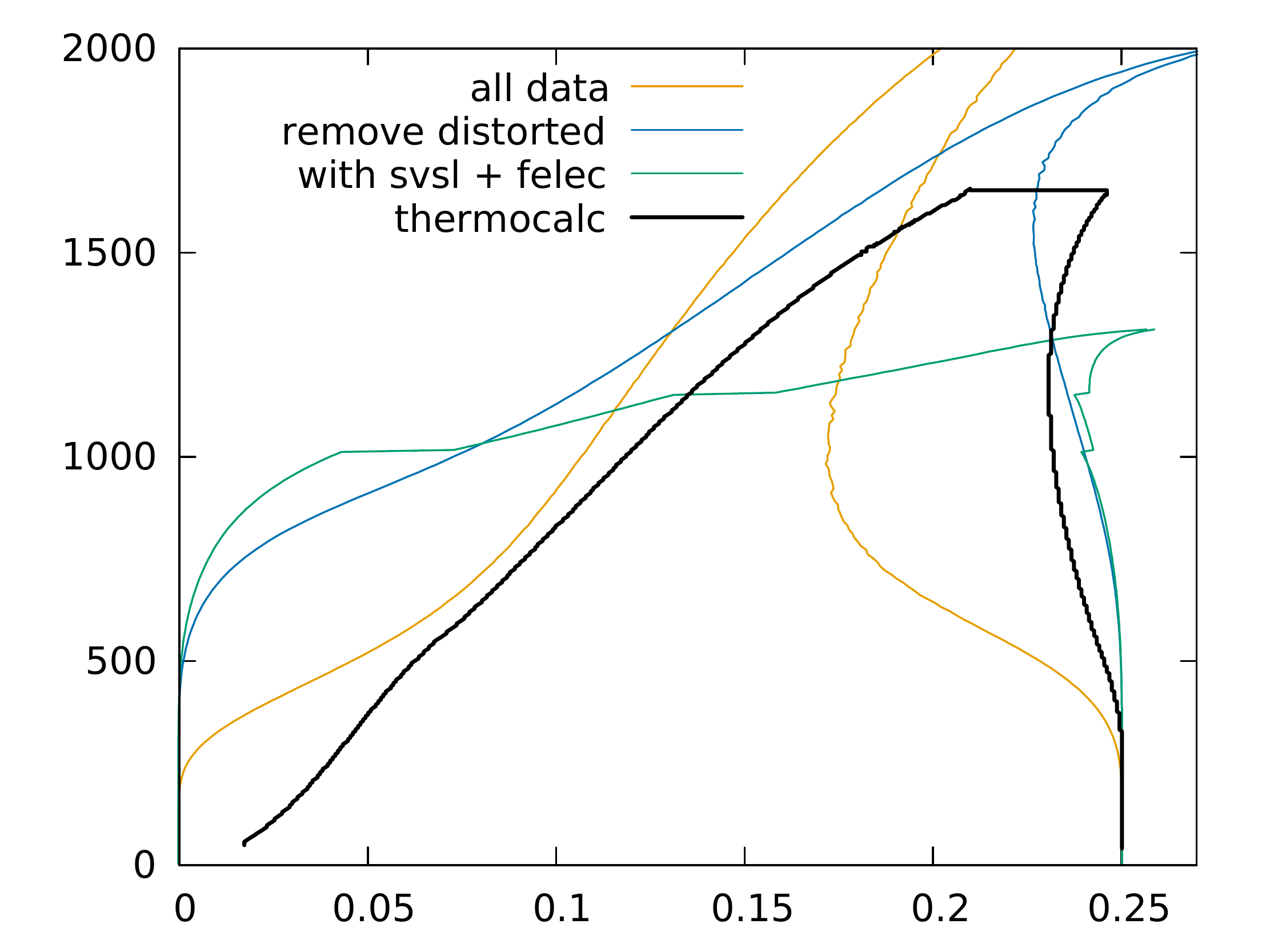} 
\caption{Phase diagram calculated with \phb using
  temperature-dependent coefficients with phonon and electron
  contributions. 
}
\label{fig:phbfinal1}
\end{figure}

Although the \phb-help page states that the ECI input file is
\texttt{eci.out}, \phb does actually read a \texttt{teci.out} if
provided. To calculate the phase diagram, we just run it:
\begin{verbatim}
phb -gs1=0 -gs2=1 -dT=10 -dx=0.001 -er=50 -keV -ltep=1.e-3 
  -o=phase01.out > phase01-run
\end{verbatim}

The result of the calculation is shown in
figure~\ref{fig:phbfinal1}. Up to 
a temperature of 1000~K, it agrees well with the previous
calculations. However, at larger temperatures, the phase separation
line looks incorrect. It might be possible that with the additional
degrees of freedom, some intermediate phases are stabilized (for
example, the jump at a concentration of about 0.125 may be due to a
\ch{Ni_7Al}-phase).  However,  to find out
exactly what is happening here is beyond the scope of this tutorial.

\section{Conclusion}
ATAT is a powerful tool to calculate phase boundaries and other
material properties. But as the experiments here have already
shown, it is not a program that can just be started and yields
publishable results without user intervention. 
Critical evaluation of
all results, experimentation, and a mixture of different approaches may
be needed to understand even a simple system like the ones we looked
at here.

\section{Acknowledgements}
Thanks to Axel van der Walle  for answering a lot of
questions on ATAT and reading the manuscript and to Jesper Kristensen
for many helful hints on ATAT.

\bibliography{fullCollection}  

\begin{thebibliography}{10}

\bibitem{Walle2002}
Axel Van~De Walle and Mark Asta.
\newblock {Self-driven lattice-model Monte Carlo simulations of alloy
  thermodynamic properties and phase diagrams}.
\newblock {\em Modelling and Simulation in Materials Science and Engineering},
  10(5):521--538, 2002.

\bibitem{Walle2009}
Axel Van~De Walle.
\newblock {Multicomponent multisublattice alloys, nonconfigurational entropy
  and other additions to the Alloy Theoretic Automated Toolkit}.
\newblock pages 1--25, 2009.

\bibitem{frenkel2001understanding}
Daan Frenkel and Berend Smit.
\newblock {\em Understanding molecular simulation: from algorithms to
  applications}, volume~1.
\newblock Elsevier, 2001.

\bibitem{Morgan2005}
Dane Morgan.
\newblock {Computational Laboratory : Monte Carlo for Phase Stability
  Calculations Atomistic Modeling Toolbox ( AMTB )}.
\newblock 2005.

\bibitem{walle}
A.~van~de Walle, M.~Asta, and G.~Ceder.
\newblock {Talk "Automatic first-principles phase diagrams calculations"}.
\newblock
  \url{https://www.brown.edu/Departments/Engineering/Labs/avdw/atat/atattalk.pdf},
  2002.

\bibitem{gnuplot}
T~Williams, C~Kelley, HB~Br{\"o}ker, J~Campbell, R~Cunningham, D~Denholm,
  E~Elber, R~Fearick, C~Grammes, and L~Hart.
\newblock Gnuplot 4.5: An interactive plotting program. 2011.
\newblock \url{http://www. gnuplot. info}, 2017.

\bibitem{Soisson2000}
F.~Soisson and G.~Martin.
\newblock {Monte Carlo simulations of the decomposition of metastable solid
  solutions: Transient and steady-state nucleation kinetics}.
\newblock {\em Physical Review B - Condensed Matter and Materials Physics},
  62(1):203--214, 2000.

\bibitem{porter2009phase}
David~A Porter, Kenneth~E Easterling, and Mohamed Sherif.
\newblock {\em Phase Transformations in Metals and Alloys, (Revised Reprint)}.
\newblock CRC press, 2009.

\bibitem{kristensen}
Jesper Kristensen.
\newblock Vasp meets atat.
\newblock
  \url{http://www.jespertoftkristensen.com/JTK/Blog/Entries/2013/11/7_Converting_str.out_to_POSCAR_and_vice_versa.html},
  2014.

\bibitem{ase}
Ask~Hjorth Larsen, Jens~J{\o}rgen Mortensen, Jakob Blomqvist, Ivano~E Castelli,
  Rune Christensen, Marcin Du{\l}ak, Jesper Friis, Michael~N Groves, Bj{\o}rk
  Hammer, Cory Hargus, et~al.
\newblock The atomic simulation environment--a python library for working with
  atoms.
\newblock {\em Journal of Physics: Condensed Matter}, 29(27):273002, 2017.

\bibitem{kristensen2}
Jesper Kristensen.
\newblock How to visualize the clusters found in atat.
\newblock
  \url{http://www.jespertoftkristensen.com/JTK/Blog/Entries/2013/11/8_How_to_visualize_the_clusters_found_in_ATAT.html},
  2014.

\bibitem{Woodward2014}
C.~Woodward, Axel Van~De Walle, Mark Asta, and D.R. Trinkle.
\newblock {First-principles study of interfacial boundaries in Ni--Ni3Al}.
\newblock {\em Acta Materialia}, 75:60--70, 2014.

\bibitem{rosler2007mechanical}
Joachim R{\"o}sler, Harald Harders, and Martin B{\"a}ker.
\newblock {\em Mechanical behaviour of engineering materials: metals, ceramics,
  polymers, and composites}.
\newblock Springer Science \& Business Media, 2007.

\bibitem{thermocalc}
Jan-Olof Andersson, Thomas Helander, Lars H{\"o}glund, Pingfang Shi, and
  Bo~Sundman.
\newblock Thermo-calc \& dictra, computational tools for materials science.
\newblock {\em Calphad}, 26(2):273--312, 2002.

\end{thebibliography}
 
\end{document}